\documentclass[fleqn,usenatbib]{mnras}

\usepackage{newtxtext,newtxmath}
\usepackage[T1]{fontenc}

\DeclareRobustCommand{\VAN}[3]{#2}
\let\VANthebibliography\thebibliography
\def\thebibliography{\DeclareRobustCommand{\VAN}[3]{##3}\VANthebibliography}

\usepackage{graphicx}	% Including figure files
\usepackage{amsmath}	% Advanced maths commands
\title[Phase structure of water-rich sub-Neptunes]{How Deep Is the Ocean? Exploring the phase structure of water-rich sub-Neptunes}

\author[M. C. Nixon and N. Madhusudhan]{
Matthew C. Nixon\thanks{E-mail: mnixon@ast.cam.ac.uk}
and Nikku Madhusudhan\thanks{E-mail: nmadhu@ast.cam.ac.uk}
\\
Institute of Astronomy, University of Cambridge, Madingley Road, Cambridge CB3 0HA, UK
}

\date{Accepted XXX. Received YYY; in original form ZZZ}

\pubyear{2020}

\begin{document}
\label{firstpage}
\pagerange{\pageref{firstpage}--\pageref{lastpage}}
\maketitle

\begin{abstract}
Understanding the internal structures of planets with a large H$_2$O component is important for the characterisation of sub-Neptune planets. The finding that the mini-Neptune K2-18b could host a liquid water ocean beneath a mostly hydrogen envelope motivates a detailed examination of the phase structures of water-rich planets. To this end, we present new internal structure models for super-Earths and mini-Neptunes that enable detailed characterisation of a planet's water component. We use our models to explore the possible phase structures of water worlds and find that a diverse range of interiors are possible, from oceans sandwiched between two layers of ice to supercritical interiors beneath steam atmospheres. We determine how the bulk properties and surface conditions of a water world affect its ocean depth, finding that oceans can be up to hundreds of times deeper than on Earth. For example, a planet with a 300$\,$K surface can possess H$_2$O oceans with depths from 30--500$\,$km, depending on its mass and composition. We also constrain the region of mass--radius space in which planets with H/He envelopes could host liquid H$_2$O, noting that the liquid phase can persist at temperatures up to 647$\,$K at high pressures of $218$--$7\times10^4\,$bar. Such H/He envelopes could contribute significantly to the planet radius while retaining liquid water at the surface, depending on the planet mass and temperature profile. Our findings highlight the exciting possibility that habitable conditions may be present on planets much larger than Earth.
\end{abstract}

\begin{keywords}
planets and satellites: interiors -- planets and satellites: composition -- planets and satellites: general
\end{keywords}

%%%%%%%%%%%%%%%%%%%%%%%%%%%%%%%%%%%%%%%%%%%%%%%%%%

%%%%%%%%%%%%%%%%% BODY OF PAPER %%%%%%%%%%%%%%%%%%

\section{Introduction}

One of the most intriguing findings of recent planetary detection missions is the ubiquity of sub-Neptune planets, whose radii lie between 1--4$R_{\oplus}$, i.e., larger than Earth but smaller than Neptune \citep{Batalha2013,Fulton2017}. These planets are often grouped into two categories, super-Earths and mini-Neptunes, distinguished by whether their radii are greater than or less than $\sim$1.75$R_{\oplus}$ \citep{Fulton2018}. With no solar system analogues, the characterisation of such planets is an important challenge in exoplanet science. In particular, understanding whether these planets are typically large rocky planets, scaled-down ice giants, or some intermediate between the two is crucial in order to place these planets in the context of the broader exoplanet population. Planets in this regime are likely to contain a substantial amount of H/He and/or H$_2$O \citep{Rogers2015,Zeng2019}, and so detailed forward modelling of planetary internal structures including these components is important for their characterisation.

Recently, \citet{Madhu2020} conducted a joint analysis of the interior and atmosphere of K2-18b, a habitable-zone mini-Netpune \citep{Foreman2015,Cloutier2017,Benneke2019,Tsiaras2019}. One of the key findings of that study was that for certain solutions, K2-18b could host a liquid water ocean beneath its hydrogen-rich atmosphere. This result provides motivation for a detailed investigation of the thermodynamic conditions and phase structures of the H$_2$O layers of super-Earths and mini-Neptunes in general. In this paper, we aim to characterise the interiors of planets with a significant H$_2$O mass fraction, both with and without H/He envelopes. We use detailed internal structure models to explore the range of phases of H$_2$O that are accessible within planetary interiors. This allows us to address a number of topics including the possible depths of liquid water oceans on such planets, and the phases of water that may be found at the interface between the H$_2$O and H/He layers of an exoplanet.

Internal structure modelling has long been used to link a planet's composition to its observable bulk properties (mass, radius and equilibrium temperature). Some of the earliest examples of these models were developed by \citet{Zapolsky1969}, who found mass--radius ($M$--$R$) relations for zero-temperature spheres made from a range of chemical species. Their methods have subsequently been developed further for exoplanets. \citet{Valencia2006} modelled Mercury to super-Earth sized planets with different core and mantle compositions incorporating thermal effects, finding that the $M$--$R$ relation differed depending on whether these planets are primarily rocky or icy. \citet{Seager2007} explored models of isothermal planets consisting of iron, silicates, water, and carbon compounds, as well as H/He, and noted a clear distinction between the radii of planets with gaseous envelopes and those without. $M$--$R$ relations for planets with sizes varying across several orders of magnitude were also computed by \citet{Fortney2007}, who combined an ice+rock interior with a H/He envelope and calculated how planetary evolution affects the interior in order to link the age of a planetary system to the internal structure of its planets.

In this paper we focus on planets which contain a substantial H$_2$O component, and which may also possess an extended H/He envelope. Since we are interested in the phase structure of the water components of these planets, it is important to accurately model the effect of temperature variations within the H$_2$O layer. A number of previous studies have considered the importance of thermal effects when modelling such planets. Early studies suggested that temperature variations did not significantly alter the $M$--$R$ relation for water worlds \citep{Sotin2007,Grasset2009}, however these works were generally limited to low-temperature planets with either liquid or icy surfaces. More recent work has suggested that thermal effects can have a sizeable impact on the $M$--$R$ relation. \citet{Madhu2015} showed that for highly irradiated H$_2$O-rich planets, the atmospheres can contribute significantly to the observed radii. \citet{Thomas2016} explored thermal effects on the $M$--$R$ relation for water worlds in more detail, focusing on planets with surfaces in liquid or supercritical phases, and finding that higher surface temperatures could lead to a large increase in radius. For example, changing the surface temperature from 300--1000$\,$K could increase the radius of a 1--10$M_{\oplus}$ planet by up to 25\%. \citet{Otegi2020} also demonstrated that changes in a planet's temperature profile can substantially alter the $M$--$R$ relation for sub-Neptunes.  \citet{Mousis2020} modelled planets with steam atmospheres and supercritical H$_2$O layers, which can inflate the radii of water worlds without invoking a H/He envelope, while \citet{Turbet2020} also found that irradiated planets could possess inflated H$_2$O layers, applying their results to the TRAPPIST-1 system.

Another important aspect is the possible extent of oceans on H$_2$O-rich planets. A number of past studies have explored this to some extent. For example, \citet{Leger2004} calculated the depths of oceans on planets with fixed mass ($6M_{\oplus}$) and an adiabatic H$_2$O layer with fixed mass fraction (50\%) across several different surface temperatures, reporting depths of 60$\,$km for $T_0=273\,$K, 72$\,$km for $T_0=280\,$K and 133$\,$km for $T_0=303\,$K. Similarly, planets with 50\% H$_2$O by mass were considered by \citet{Sotin2007}, however in this case they used an isothermal temperature profile and the surface temperature was fixed to 300$\,$K while the planet mass was varied. They found that a $1M_{\oplus}$ planet should have an ocean that is 150$\,$km deep, decreasing to 50$\,$km for a $10M_{\oplus}$ planet. The decrease in ocean depth with increasing mass is a result of the higher surface gravity of more massive planets with the same composition. \citet{Alibert2014} considered the limiting case where a planet has the maximum amount of H$_2$O possible while avoiding a high-pressure ice layer. In this scenario, the total H$_2$O mass remains approximately constant (at $\sim$0.03$M_{\oplus}$) as the planet mass changes. A study by \citet{Noack2016} noted that the maximum possible ocean depth for a given planet varies with its mass, composition and surface temperature. At 300$\,$K their results agree with \citet{Sotin2007}, but they found that between 290 and 370$\,$K, a 10$\,$K increase in surface temperature leads to a 14--16\% increase in ocean depth. 

More broadly, understanding a planet's phase structure can provide insight into its bulk geophysical properties, as has been demonstrated for icy moons in the Solar System \citep[e.g. ][]{Hsu2015,Soderlund2020}. The presence or absence of a liquid water layer has important consequences for planetary habitability \citep{Lammer2009}. Other works have explored the general phase structures of H$_2$O-rich planets. For example, \citet{Zeng2014} studied the temperature evolution of the interiors of water-rich planets. They found that the phase structures of these planets may change as the planet cools, and that planets older than $\sim$3$\,$Gyr should have mostly solid H$_2$O layers, assuming they are not highly irradiated. 

As well as H$_2$O-rich planets without an extended atmosphere, we are also interested in the interiors of planets smaller than Neptune which retain an extended H/He envelope. Only a small amount of H/He is required to have a large impact on planetary radius \citep{Lopez2014}. \citet{Rogers2011} showed that increasing the equilibrium temperature of such planets also inflates their radii significantly, with the effect being more pronounced for lower-mass planets. The extent of such an envelope as well as its temperature structure determines the thermodynamic conditions at the boundary between the envelope and the remaining interior. This is a crucial factor in determining the internal structure and surface conditions of a water layer. Various approaches have been taken when incorporating hydrogen-rich envelopes into internal structure models, ranging from isothermal H/He layers \citep{Seager2007} to envelopes following an analytic temperature profile \citep{Rogers2011,Valencia2013} or a temperature profile derived from self-consistent atmospheric modelling \citep{Madhu2020}.

Internal structure models have been used to characterise observed super-Earths and mini-Neptunes, both through population studies and application to specific planets. At the population level, \citet{Rogers2015} used $M$--$R$ relations along with a hierarchical Bayesian analysis of the \textit{Kepler} sample of planets to show that planets with radii $\gtrsim 1.6 R_{\oplus}$ are most likely volatile-rich. \citeauthor{Dorn2015} (\citeyear{Dorn2015,Dorn2017}) developed a Bayesian framework to infer super-Earth compositions and place constraints on the extent of a volatile envelope, finding that solutions are often highly degenerate, with a range of compositions able to explain a given mass and radius. \citet{Lopez2012} predicted that the H$_2$O component of planets in the Kepler-11 system would be in vapour, molecular fluid or ionic fluid phases. As one of the first exoplanets smaller than Neptune to have both a measured mass and radius, the planet GJ~1214b has been the subject of numerous studies aiming to characterise its interior. The bulk properties of the planet suggest the presence of a H/He envelope, with the possibility of a H$_2$O layer in the interior. \citet{Rogers2010b} showed that the planet most likely has a substantial H/He layer above a layer of H$_2$O, which would be in supercritical and sublimated ice phases. The presence of a gaseous layer was also suggested by \citet{Nettelmann2011}, whose models favoured a metal-enriched H/He atmosphere with a significant H$_2$O mass fraction. An upper limit on the H/He mass fraction of 7\% was proposed by \citet{Valencia2013}. As mentioned previously, \citet{Madhu2020} analysed the interior and atmosphere of K2-18b, a habitable-zone temperature mini-Neptune, constraining the planet's H/He mass fraction to $\lesssim 6\%$. Constraints on the planet's interior structure showed that conditions at the surface of the H$_2$O layer could range from supercritical to liquid phases.

In this paper we present internal structure models for super-Earths and mini-Neptunes, with the aim of thoroughly exploring the phase structures of H$_2$O layers on such planets. We describe our model in Section \ref{section:methods}, including our compilation of the H$_2$O equation of state (EOS) that is valid across a wide pressure and temperature range, and approaches to incorporating the temperature structure of a H/He envelope. In Section \ref{section:validation} we validate our model against previously published results. We then use our models to explore the phase structures of H$_2$O-rich planets in Section \ref{section:results}. We investigate in detail how the ocean depth depends on observable properties such as surface gravity and temperature. We also constrain the range of masses and radii of planets that might host liquid water beneath H/He envelopes. We explore the wide range of internal structures that may be present on H$_2$O-rich planets with different surface conditions. We also show $M$--$R$ relations for planets with mixed H/He-H$_2$O envelopes. Finally in Section \ref{section:discussion} we summarise our results and discuss possible caveats, implications of our findings and avenues for future study.

\begin{figure}
\includegraphics[width=\columnwidth,trim={0 1.6cm 0 0},clip]{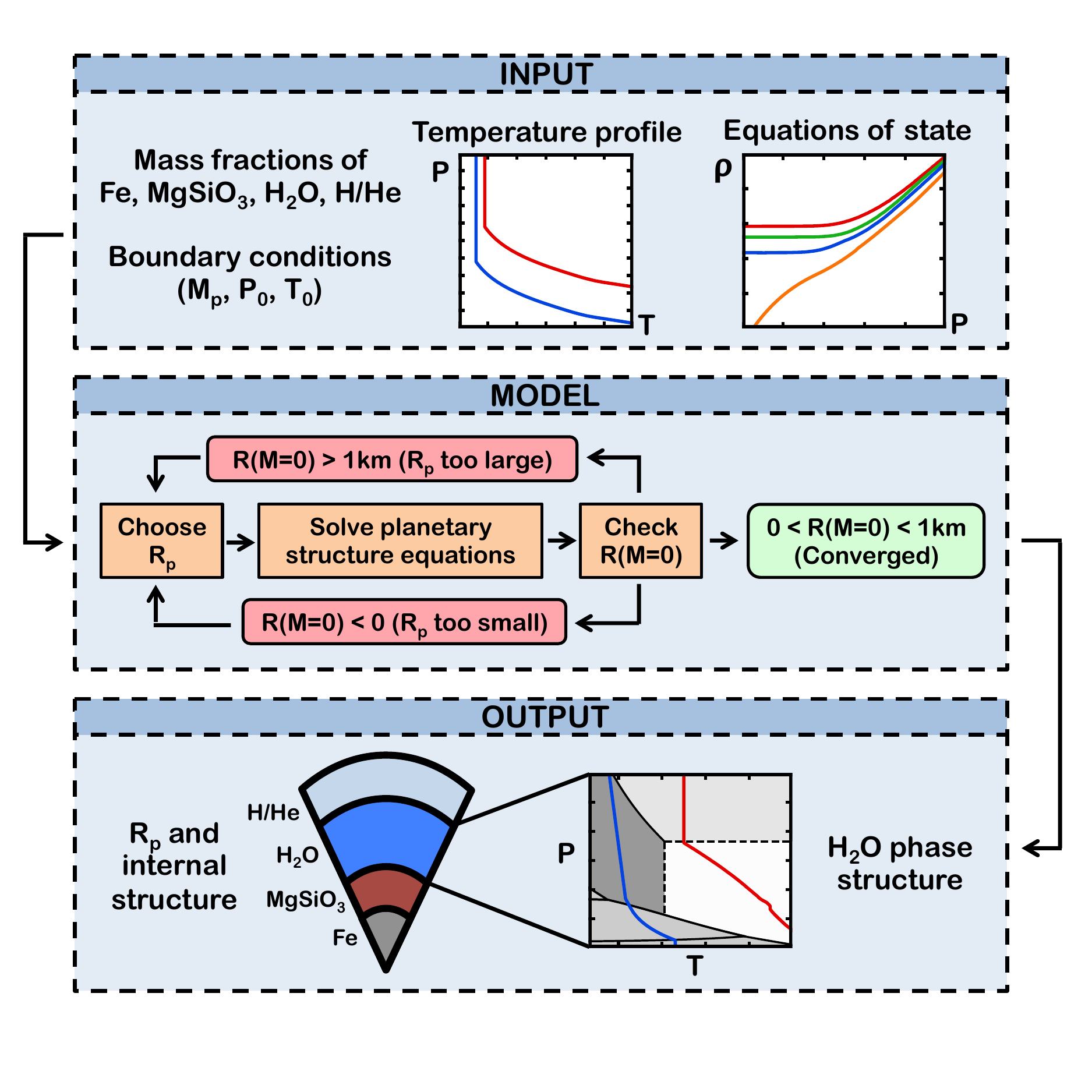}
    \caption{Architecture of the internal structure model used in this study.}
    \label{fig:arch}
\end{figure}

%%%%%%%%%%%%%%%%%%%%%%%%%%%%%%%%%%%%%%%%%%%%%%%%%%%%%%%%%%%%%%%%%%%%%%%%%%

\section{Methods} \label{section:methods}

%%%%%%%%%%%%%%%%%%%%%%%%%%%%%%%%%%%%%%%%%%%%%%%%%%%%%%%%%%%%%%%%%%%%%%%%%%

The canonical planetary internal structure model consists of a differentiated two-component nucleus made up of an iron core and silicate mantle, a layer of H$_{2}$O and/or a H/He envelope \citep[e.g.,][]{Leger2004,Fortney2007,Seager2007,Sotin2007,Valencia2007}. We adopt this approach and follow the standard assumption that the core and mantle are fully differentiated. However, since it has been suggested that water and hydrogen are miscible in the interiors of giant planets \citep{Soubiran2015}, we therefore allow for a mixed H/He and H$_2$O envelope as well as a differentiated structure with an outer H/He envelope on top of the H$_2$O layer. The components of our model are depicted in Figure \ref{fig:arch} and described in detail in this section.

\subsection{Planetary structure equations}

Our model solves the equations which determine the interior structure of a planet: the mass continuity equation,
\begin{equation}
    \frac{dR}{dM} = \frac{1}{4 \pi R^2 \rho},
\end{equation}
where $M$ is the mass of a spherical shell of material internal to a radius $R$ and density $\rho$, and the equation of hydrostatic equilibrium,
\begin{equation}
    \label{eq:hydro_eqm}
    \frac{dP}{dM} = - \frac{GM}{4\pi R^4},
\end{equation}
where $P$ is the pressure at the shell. Linking these equations requires an equation of state (EOS)  $\rho = \rho(P,T)$ as well as a pressure-temperature ($P$--$T$) profile $T = T(P)$, or simply $\rho = \rho(P)$ for a temperature-independent EOS. The EOS for each component is described in Section \ref{subsection:eos} and the temperature profiles used in this study are discussed in Section \ref{subsection:tp}.

These equations are solved using a fourth-order Runge-Kutta scheme, with the mass interior to a shell taken as the independent variable. Previous internal structure models have performed the integration either by starting at the surface of the planet and proceeding inward \citep[e.g.,][]{Rogers2010a,Madhu2012,Thomas2016,Madhu2020}, or by integrating outward from the centre \citep[e.g.,][]{Seager2007,Sotin2007,Noack2016}. We choose to integrate inward from the surface since in this case the boundary conditions to be specified are surface conditions of the planet, which are more closely linked to observable parameters than the conditions at the centre of the planet. For example, \citet{Madhu2020} used the retrieved reference pressure from the transmission spectrum of K2-18b as a boundary condition for the interior models. Our boundary conditions are therefore the temperature and pressure at the photosphere of the planet (i.e. the conditions at $R=R_p$.)

We solve for the planetary radius $R_p$ at a given mass $M_p$ and mass fractions $x_i = M_i/M_p$ of iron, silicates, H$_2$O and H/He. $R_p$ is found using a bisection root-finding scheme. The value of $R_p$ is updated iteratively for solving the structure equations until the conditions $0 < R(M=0) < 1\,$km are satisfied.

\subsection{Equations of state} \label{subsection:eos}

\begin{table}
	\centering
	\caption{Parameters for the EOS of Fe ($\epsilon$) from \citet{Anderson2001} and MgSiO$_3$ (perovskite) from \citet{Karki2000}.}
	\hfill \\
	\label{tab:eos_params}
	\setlength{\arrayrulewidth}{1.3pt}
	\begin{tabular}{ccccc}
		\hline
		Component & $B_0$ (GPa) & $B'_0$ & $B''_0$ (GPa$^{-1}$) & $\rho_0$ (kg m$^{-3}$) \\
		\hline
		Fe & 156.2 & 6.08 & n/a & 8300 \\
		MgSiO$_3$ & 247 & 3.97 & -0.016 & 4100 \\
		\hline
	\end{tabular}
\end{table}

\begin{figure}
\includegraphics[width=\columnwidth]{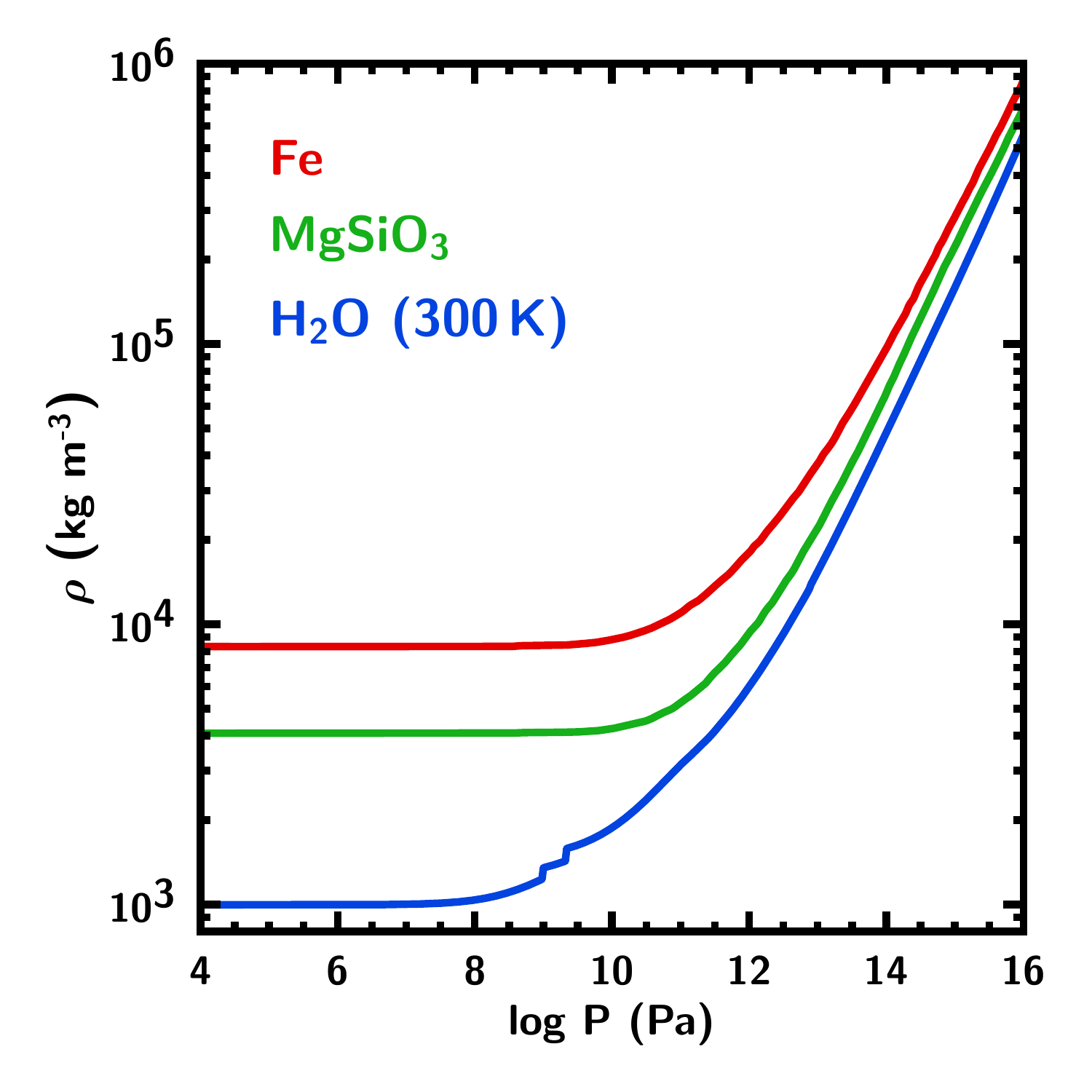}
    \caption{Equations of State for Fe and MgSiO$_3$ used in our model. We use a temperature-indpendent EOS adopted from \citet{Seager2007} for each of these components. The EOS for H$_2$O at 300 K is also shown for comparison.}
    \label{fig:fe_si_eos}
\end{figure}

\begin{figure*}
\includegraphics[width=\textwidth]{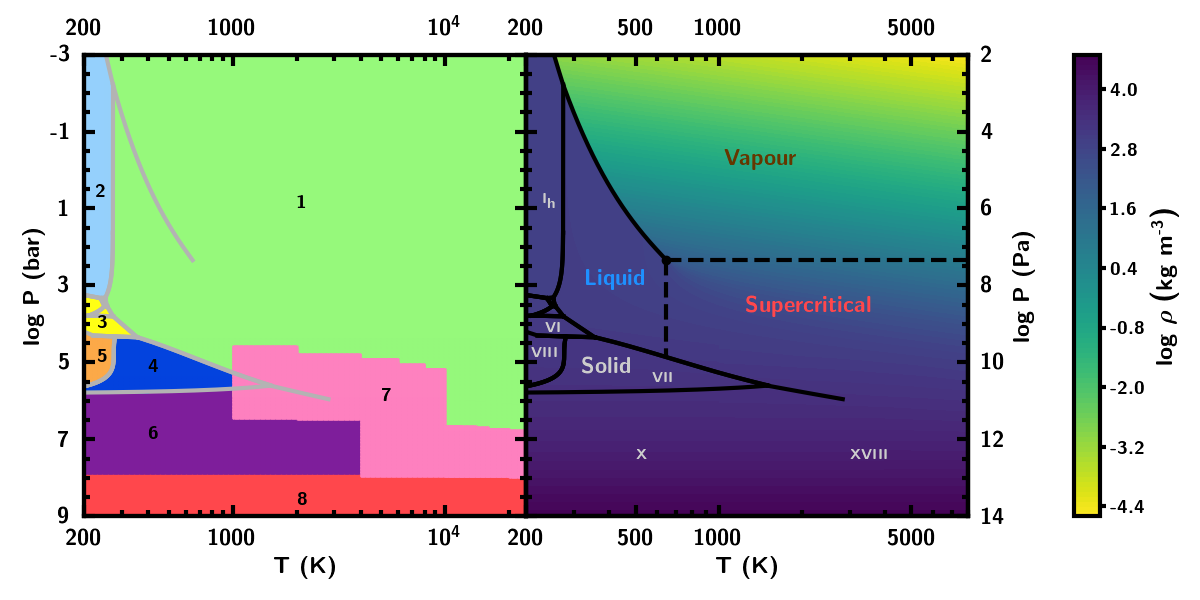}
    \caption{\textit{Left:} Data sources for the H$_2$O EOS used in our model: 1. \citet{Wagner2002}. 2. \citet{Feistel2006}. 3. \citet{Journaux2020_EOS}. 4. \citet{Fei1993}. 5. \citet{Klotz2017}, \citet{Fei1993}. 6. \citet{Seager2007}. 7. \citet{French2009}. 8. \citet{Salpeter1967}. The EOS of \citet{Salpeter1967} is also used for $P>10^{14}\,$Pa. \textit{Right:} Phase diagram of H$_2$O \citep{Wagner2002,Dunaeva2010}. The contour plot shows the EOS  $\rho = \rho(P,T)$ used in our model. Regions of $P$--$T$ space are labelled with their corresponding phase. The transition from ice X to ice XVIII (also called superionic ice) occurs at approximately 2000$\,$K \citep{Millot2019}.}
    \label{fig:h2o_eos}
\end{figure*}

\begin{figure*}
\includegraphics[width=\textwidth]{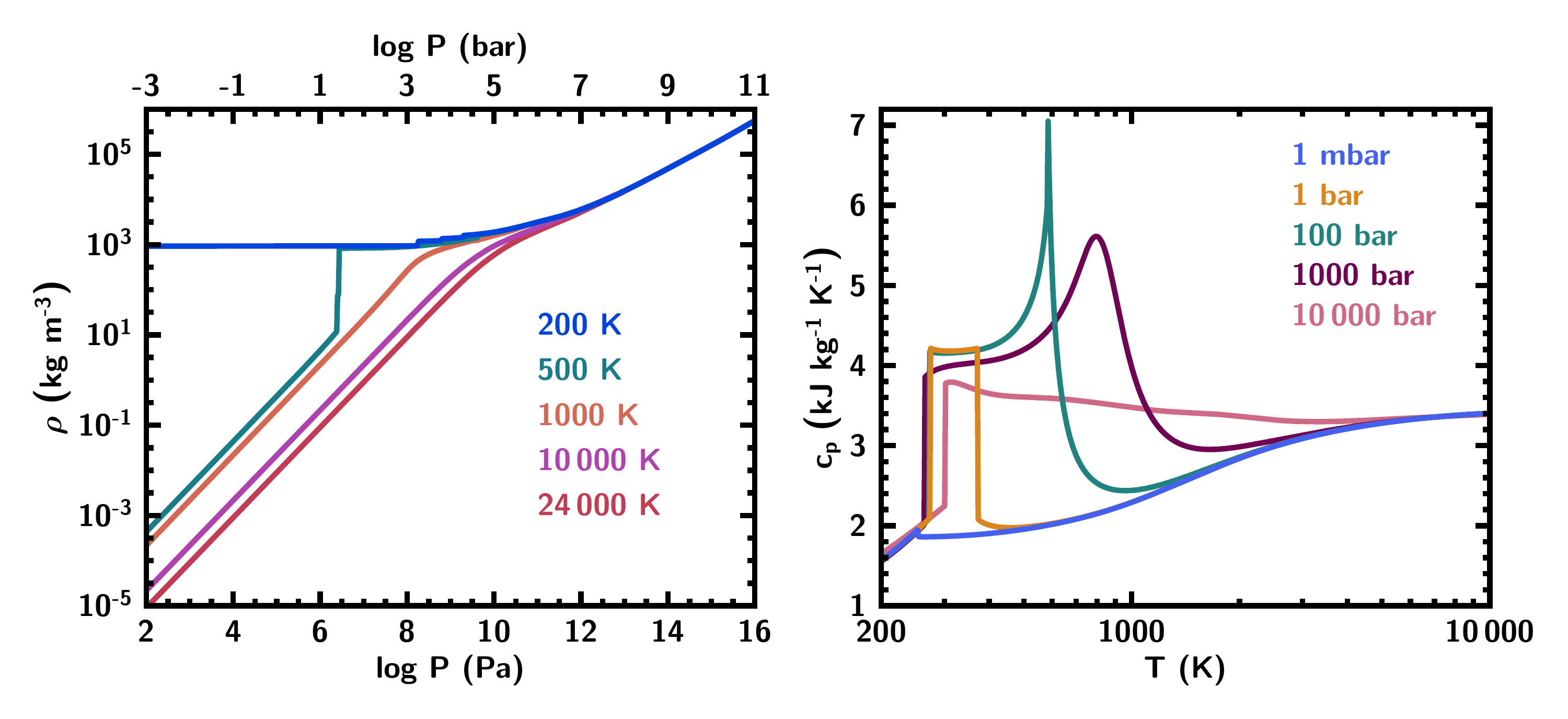}
    \caption{\textit{Left:} Density of H$_2$O as a function of pressure for several different isotherms. At 500$\,$K there is a phase transition from vapour to liquid at 26.4$\,$bar. \textit{Right:} Specific heat capacity $c_{\rm p}$ of H$_2$O as a function of temperature for a range of isobars. The value of $c_{\rm p}$ increases sharply across the ice-liquid phase transition and decreases across the liquid-vapour boundary. At high temperatures away from the phase boundaries $c_{\rm p}$ does not vary significantly.}
    \label{fig:h2o_rho_cp}
\end{figure*}

\begin{figure}
\includegraphics[width=\columnwidth]{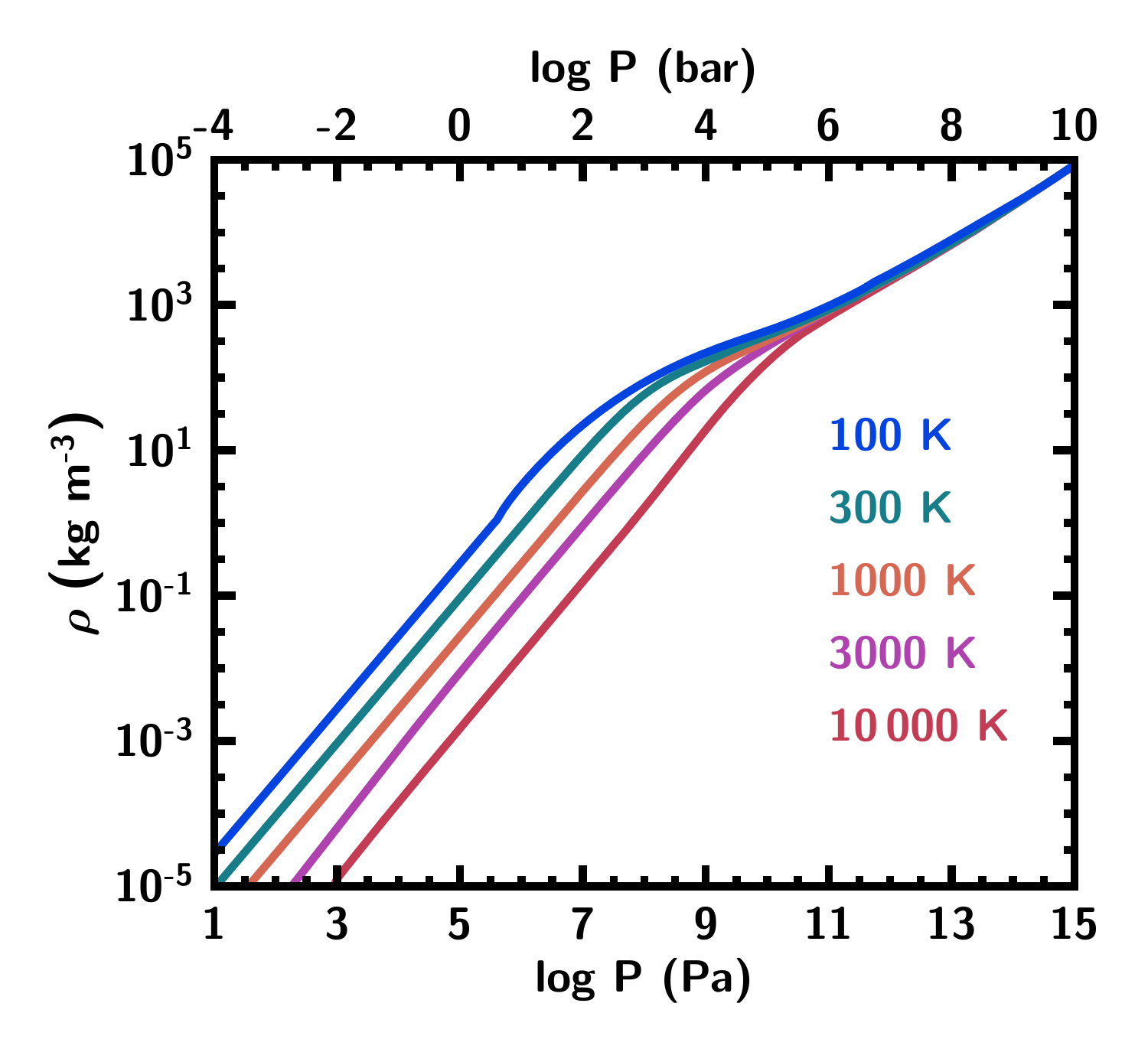}
    \caption{EOS for H/He used in our model for different isotherms. At low pressures ($\lesssim 10^7\,$Pa) the EOS behaves like an ideal gas, and at high pressures ($\gtrsim\,10^{12}\,$Pa) it becomes temperature-independent. The EOS data is taken from \citet{Chabrier2019}.}
    \label{fig:h_he_eos}
\end{figure}

Previous works have explored a number of EOS prescriptions for different model components. Some early studies used an isothermal EOS for each component \citep{Zapolsky1969,Seager2007}, while others considered temperature dependence throughout the interior \citep{Valencia2006,Sotin2007}. It has been demonstrated that thermal effects in the iron and silicate layers do not substantially alter the $M$--$R$ relation \citep{Grasset2009,Howe2014}, and therefore several models treat the inner layers as isothermal with a temperature-dependent prescription for other components \citep[e.g.,][]{Rogers2010a,Zeng2013,Thomas2016}. Although a number of past models have used isothermal H$_2$O layers \citep[e.g.,][]{Hubbard1989,Seager2007}, several works have shown that thermal effects in this layer can significantly affect the $M$--$R$ relation \citep{Thomas2016,Mousis2020,Turbet2020}. Constructing an EOS for H$_2$O can be challenging due to the need to accurately describe the behaviour of the many different phases of water. For example, the International Association for the Properties of Water and Steam (IAPWS) provide a detailed functional EOS \citep{Wagner2002} covering liquid, vapour and some of the supercritical phase, and \citet{French2009} present an EOS that is valid for high-pressure ice. One way of avoiding this problem is to compile a patchwork EOS that uses different prescriptions for different phases. This approach was taken by \citet{Thomas2016}, who compiled a temperature-dependent H$_2$O EOS valid at pressures from $10^5-10^{22}\,$Pa and temperatures from $300-24\,000\,$K). A similar method was adopted by \citet{Mazevet2019}, who constructed an EOS covering the liquid, gas and superionic regimes.

For the iron and silicate layers we adopt the isothermal (room-temperature) EOS described in \citet{Seager2007}. We present a revised and updated version of the temperature-dependent H$_2$O EOS from \citet{Thomas2016} and we use the recently published H/He EOS from \citet{Chabrier2019}. We now describe each of these prescriptions in detail.

\subsubsection{Iron}

Figure \ref{fig:fe_si_eos} shows the temperature-independent EOS that we use for each component of the nucleus, alongside our H$_2$O EOS at 300$\,$K for comparison. We adopt the EOS of the $\epsilon$ phase of Fe from \citet{Seager2007}, which consists of a Vinet fit \citep{Vinet1989} to experimental data from \citet{Anderson2001} at lower pressures ($P < 2.09 \times 10^{13}\,$Pa), and transitions to the Thomas-Fermi-Dirac (TFD) EOS as described in \citet{Salpeter1967} at higher pressures. The Vinet EOS takes the functional form

\begin{equation}
    P = 3 B_0 \eta^{2/3} \left(1 - \eta^{-1/3} \right) \exp \left[ \frac{3}{2} \left( B'_0 -1 \right) \left(1-\eta^{-1/3} \right) \right]
\end{equation}
where $B_0 = \rho(\partial P / \partial \rho)|_T$ is the isothermal bulk modulus, $B'_0$ is the pressure derivative of $B_0$ and $\eta = \rho/\rho_0$, where $\rho_0$ is the ambient density. The values used in this study for those quantities are shown in Table \ref{tab:eos_params}. For the TFD EOS, electrons are treated as a gas of non-interacting particles in a slowly-varying potential. At lower pressures chemical bonds and the crystal structure of a material play an  important role in determining the EOS, and the TFD approximation breaks down since these effects are not considered. However at very high pressures, when kinetic energy dominates over Coulomb energy, TFD theory can yield an accurate EOS. The low-pressure Fe EOS smoothly transitions into the TFD EOS allowing the pressure range to be extended beyond what is obtainable solely from experimental data.

\subsubsection{Silicates}

The silicate EOS used in our model is also adopted from \citet{Seager2007} and comes from the perovskite phase of MgSiO$_3$. For $P < 1.35 \times 10^{13}\,$Pa this takes the form of a fourth-order finite strain Birch-Murnaghan EOS \citep{Birch1952} fit to a density functional calculation from \citet{Karki2000}. The fourth-order Birch-Murnaghan equation is

\begin{equation}
\begin{split}
    P & = \frac{3}{2} B_0 \left(\eta^{7/3} - \eta^{5/3} \right) \bigg\{ 1 + \frac{3}{4} \left( B'_0 -4 \right) \left(\eta^{2/3} -1 \right) \\ 
    & + \frac{3}{8} B_0 \left( \eta^{2/3} -1 \right)^2 \left[ B_0 B''_0 + B'_0 \left(B'_0 - 7 \right) + \frac{143}{9} \right] \bigg\}
\end{split}
\end{equation}
where $B''_0$ is the second pressure derivative of $B_0$. The values used in this study for the relevant quantities are shown in Table \ref{tab:eos_params}. As with the Fe EOS, this smoothly transitions into a TFD EOS at high pressures.

\subsubsection{Water}

For the liquid, vapour and supercritical phases of H$_2$O we use the prescription presented in \citet{Thomas2016}. However, we have used different sources for several ice phases, both to extend the validity of the EOS to lower pressures and temperatures and to incorporate newer data. The resulting EOS is valid for pressures from $10^2-10^{22}\,$Pa and temperatures from $200-24\,000\,$K. Figure \ref{fig:h2o_eos} shows the various sources used to compile the EOS and their regions of validity in $P$--$T$ space. We now describe our choice of EOS for each of the relevant phases of H$_2$O.

\textit{Liquid and vapour.} We use the functional EOS from the International Association for the Properties of Water and Steam \citep[IAPWS,][]{Wagner2002}. The IAPWS EOS has been validated by numerous experiments and covers the region of $P$--$T$ space above the melting curve of H$_2$O (lowest temperature 251.2 K).

\textit{Supercritical.} \citet{Wagner2002} showed that the IAPWS formulation can be extrapolated to pressures and temperatures beyond the critical point of H$_2$O. We therefore adopt their formulation for some of the supercritical phase. However, for $T>1000\,$K and $P > 1.86 \times 10^9\,$Pa we smoothly transition to the EOS presented by \citet{French2009}. This EOS is derived from quantum molecular dynamics simulations of high-pressure ice, supercritical and superionic H$_2$O, and has been validated by experiments \citep{Knudson2012}.

\textit{Low-pressure ice.} We incorporate the EOS for ice Ih from \citet{Feistel2006} which is widely considered to be the best available formulation for this phase \citep[see e.g.][]{Journaux2020}. For ices II, III, V and VI we use the latest available EOS from \citet{Journaux2020_EOS} which is derived from experiments conducted at a range of pressures and temperatures.

\textit{High-pressure ice.} We use the \citet{French2009} EOS where applicable, covering parts of the ice VII, X, and XVIII phases. Ice XVIII exists at $P>10^{11}\,$Pa, $T>2000\,$K and is also called superionic ice \citep{Millot2019}. For the remainder of the ice VII phase we follow the approach of \citet{Fei1993}, who used a Vinet EOS with a thermal correction. \citet{Klotz2017} determined a functional form of the coefficient of volumetric thermal expansion, $\alpha$, that allows for the extrapolation of the ice VII EOS down to the ice VIII phase at lower temperatures, finding good agreement with experimental data. We therefore use their prescription for $\alpha$ to calculate an EOS for ice VIII. Thermal effects become negligible for very high pressures, and so we switch to the temperature-independent TFD theoretical EOS as described in \citet{Salpeter1967} for pressures above $7.686 \times 10^{12}\,$Pa. For intermediate regions not covered by another data source we use the EOS from \citet{Seager2007} in order to smoothly transition to the TFD regime.

\subsubsection{Hydrogen/Helium}

We use the temperature-dependent H/He EOS from \citet{Chabrier2019} for a solar helium mass fraction ($Y=0.275$), which covers pressures from 1--10$^{22}\,$Pa and temperatures from 100--10$^8\,$K. At temperatures relevant to our model, the hydrogen EOS is a combination of the semi-analytical model from \citet{Saumon1995} at low densities ($\rho \leq 50\,$kg m$^{-3}$), the model based on ab initio electronic structure calculations from \citet{Caillabet2011} at intermediate densities ($300 < \rho \leq 5000\,$kg m$^{-3}$), and the model for fully ionised hydrogen from \citet{Chabrier1998} at high densities ($\rho > 10^4\,$kg$\,$m$^{-3}$). Similarly, the helium EOS is derived using a combination of models from \citet{Saumon1995} for $\rho \leq 100\,$kg$\,$m$^{-3}$ and \citet{Chabrier1998} for $\rho > 10^5\,$kg$\,$m$^{-3}$, and ab initio calculations based on quantum molecular dynamics for $1000 < \rho \leq 10^5\,$kg$\,$m$^{-3}$. In both cases a bicubic spline procedure is used to interpolate the thermodynamic quantities between the given regimes.

The combined H/He EOS is produced using an additive volume law, which takes the form

\begin{equation}
    \frac{1}{\rho_{\rm mix}(P,T)} = \sum_i \frac{x_i}{\rho_{i}(P,T)},
\end{equation}
where $x_i$ is the mass fraction of the $i^{\rm th}$ component. This prescription does not consider interactions between the two species, but \citet{Chabrier2019} claim that the correction to the EOS from taking this into account should only be of the order of a few per cent. The resulting EOS at several different temperatures is shown in Figure \ref{fig:h_he_eos}. The same additive volume law is used to compute the density of a mixed envelope consisting of H/He and H$_2$O in this study.

\subsection{Temperature profiles} \label{subsection:tp}

The inclusion of EOS data across a wide range of pressures and temperatures enables us to consider any reasonable temperature profile within the H$_2$O and H/He layers. Deep within the interior of the planet we expect convection to dominate energy transport, leading to an adiabatic temperature profile with constant specific entropy $S$. However, at lower pressures near the planetary surface an adiabatic temperature profile may no longer be appropriate. Previous studies have used different methods to model the temperature structure of the outer envelope: for example, \citet{Fortney2007} took $P$--$T$ profiles from a grid of atmospheric models, while \citet{Rogers2011} and \citet{Valencia2013} incorporated the analytic atmospheric model from \citet{Guillot2010}. Recent works have also coupled interior models of water worlds to steam atmosphere models \citep{Mousis2020,Turbet2020}. Here we describe some of the most common approaches used to model the temperature structures of planetary interiors: analytic models and self-consistent models. We subsequently describe the approach that we use throughout most of this paper, in which the temperature profile consists of an isothermal layer above an adiabatic layer. While our model has the capability to incorporate both analytic $P$--$T$ profiles as well as those produced using a self-consistent atmospheric model, we find that isothermal/adiabatic temperature profiles are the most appropriate for this study, as we explain in Section \ref{subsec:choice}.

\subsubsection{Analytic profiles}
The temperature profile of a planet's outer envelope can be calculated using an analytic model, such as the one described in \citet{Guillot2010}, which takes the form

\begin{equation}
\begin{split}
    T^4 & = \frac{3T_{\rm int}^4}{4} \left[ \frac{2}{3} + \tau \right] \\ 
    & + \frac{3T_{\rm irr}^4}{4} f \left[ \frac{2}{3} + \frac{1}{\gamma \sqrt{3}} + \left( \frac{\gamma}{\sqrt{3}} - \frac{1}{\gamma \sqrt{3}} \right) \exp \left( -\gamma \tau \sqrt{3} \right) \right]
\end{split}
\label{eq:Guillot}
\end{equation}
which is a solution to the equations of radiative transfer assuming a grey atmosphere and the two-stream approximation. The irradiation temperature $T_{\rm irr} = f^{-1/4}T_{\rm eq}$ characterises the irradiation intensity from the host star and is related to the planetary equilibrium temperature via the redistribution factor $f$. The intrinsic temperature $T_{\rm int}$ characterises the planetary intrinsic heat flux. The ratio of visible to thermal opacities is represented by $\gamma$. This approach requires the inclusion of another differential equation in the model to solve for the optical depth $\tau$:

\begin{equation}
    \frac{d\tau}{dM} = - \frac{\kappa}{4 \pi R^2},
\end{equation}
where the opacity $\kappa$ can be specified as a function of $P$ and $T$ using, for example, the tabulated values of \citet{Freedman2008} for H/He. The analytic model described here has some limitations: for example, there is no treatment of clouds, which can have a significant impact on the form of the temperature profile \citep[e.g.,][]{Kitzmann2010} and may be prevalent in super-Earth atmospheres \citep{Kreidberg2014}. In this study, we only use analytic temperature profiles in order to validate our model against previous work that used this prescription (see Section \ref{section:validation}).

\subsubsection{Self-consistent profiles}
Another approach is to use a temperature profile that has been calculated using self-consistent atmospheric modelling \citep[e.g.,][]{Gandhi2017,Malik2019,Piette2020}. Self-consistent atmospheric models solve the equations of radiative transfer numerically under the assumptions of hydrostatic, radiative-convective and thermal equilibrium. These models are able to account for many more phenomena than the analytic prescription, such as atmospheric dynamics, clouds and particle scattering. Detailed self-consistent atmospheric modelling is not incorporated directly into our internal structure model, but temperature profiles calculated in this way can be used to obtain the density profile for the outer layers of a planet \citep{Madhu2020}. Self-consistent atmospheric modelling requires many planet-specific parameters and is more time-consuming than the other approaches discussed here, so while this method is useful for exploring the structure of a particular planet, it is less well-suited to theoretical calculations across a wide range of masses and radii. We show in Section \ref{subsubsec:ia} that an isothermal/adiabatic profile can be used in place of a self-consistently modelled profile with little change to the $M$--$R$ relation.

\subsubsection{Isothermal/adiabatic profiles} \label{subsubsec:ia}
This $P$--$T$ profile consists of an isotherm at the photospheric temperature $T_0$ down to the radiative-convective boundary, at which point the temperature profile becomes adiabatic. This approach to calculating the temperature profile allows for a high degree of flexibility while remaining simple to compute. The pressure at the radiative-convective boundary ($P_{\rm rc}$) is a free parameter.

The adiabatic temperature gradient is
\begin{equation}
    \frac{\partial T}{\partial P} \bigg|_S = \frac{\alpha T}{\rho c_{\rm p}},
\end{equation}
where $c_{\rm p}$ is the specific heat capacity at constant pressure and $\alpha$ is the coefficient of volumetric thermal expansion. \citet{Chabrier2019} present the adiabatic gradient for H/He along with their EOS and so we incorporate this directly into our model. For H$_2$O we require prescriptions for $c_{\rm p}$ and $\alpha$. \citet{Thomas2016_thesis} incorporated $c_{\rm p}$ data for the liquid and vapour phases from \citep{Wagner2002} and extrapolated this to cover all other phases of H$_2$O. We also use this data for liquid and vapour, but we do not extrapolate beyond these regions. Instead, we add data from \citet{Feistel2006} for ice Ih, \citet{Journaux2020_EOS} for ices II, III, V and VI, \citet{Fei1993} for ices VII and VIII and \citet{French2009} for the ice VII-X transition. The behaviour of $c_{\rm p}$ is summarised in the right-hand panel of Figure \ref{fig:h2o_rho_cp}. For higher pressures where $c_{\rm p}$ data is unavailable (i.e. sources 6 and 8 in Figure \ref{fig:h2o_eos}), we assume that $c_{\rm p}$ is equal to its value at the nearest point in $P$--$T$ space with available data. The true value of $c_{\rm p}$ is not required here since the EOS used for these pressures is not temperature-dependent.

We calculate $\alpha$ directly from our EOS:
\begin{equation}
    \alpha = \frac{1}{V} \frac{\partial V}{\partial T} \bigg|_P = - \frac{\partial \ln \rho}{\partial T} \bigg|_P.
\end{equation}
Transitions between different phases of H$_2$O can lead to significant discontinuities in the EOS, causing $\alpha$ to become undefined at phase boundaries. In order to avoid this, the derivative is calculated separately for each phase and smoothly interpolated across the boundary, yielding adiabats that remain continuous.

For mixed envelopes consisting of both H/He and H$_2$O, the adiabatic gradient is calculated by linear interpolation using the following formula:

\begin{equation}
    \left( \frac{\partial \log T}{\partial \log P}\bigg|_S \right)_{\rm mix} = - \frac{\Sigma_i x_i S_i \frac{\partial \log S_i}{\partial \log P}\big|_T}{\Sigma_i x_i S_i \frac{\partial \log S_i}{\partial \log T}\big|_P},
\end{equation}
with values of the specific entropy of H/He and H$_2$O taken from the same sources as the values of $c_{\rm p}$.

\subsection{Choice of temperature profiles in this study}
\label{subsec:choice}

In the rest of this paper we use isothermal/adiabatic $P$--$T$ profiles as described in the previous subsection. Here we show that these $P$--$T$ profiles are a reasonable approximation to those generated by self-consistent models. We take a number of $P$--$T$ profiles generated using the self-consistent model \textsc{genesis} \citep{Gandhi2017}, which was recently updated to model atmospheres of sub-Neptunes \citep{Piette2020}. We consider a number of models with a wide range of internal energies, which are determined by $T_{\rm int}$. $T_{\rm int}$ can be calculated using evolutionary models. For example, \citet{Lopez2014} find that, for a low-mass 5~Gyr-old planet, $T_{\rm int}$ can be as low as $\sim$10$\,$K. Conversely, \citet{Valencia2013} find that for the mini-Neptune GJ~1214b, $T_{\rm int}$ may be up to 80~K at an age of 0.1~Gyr, and \citet{Morley2017_GJ436} consider even higher values for the Neptune-mass GJ~436b, whose interior may be warmed by tidal heating. We therefore explore a range of $10$--$150\,$K for $T_{\rm int}$.

\begin{figure}
\includegraphics[width=\columnwidth]{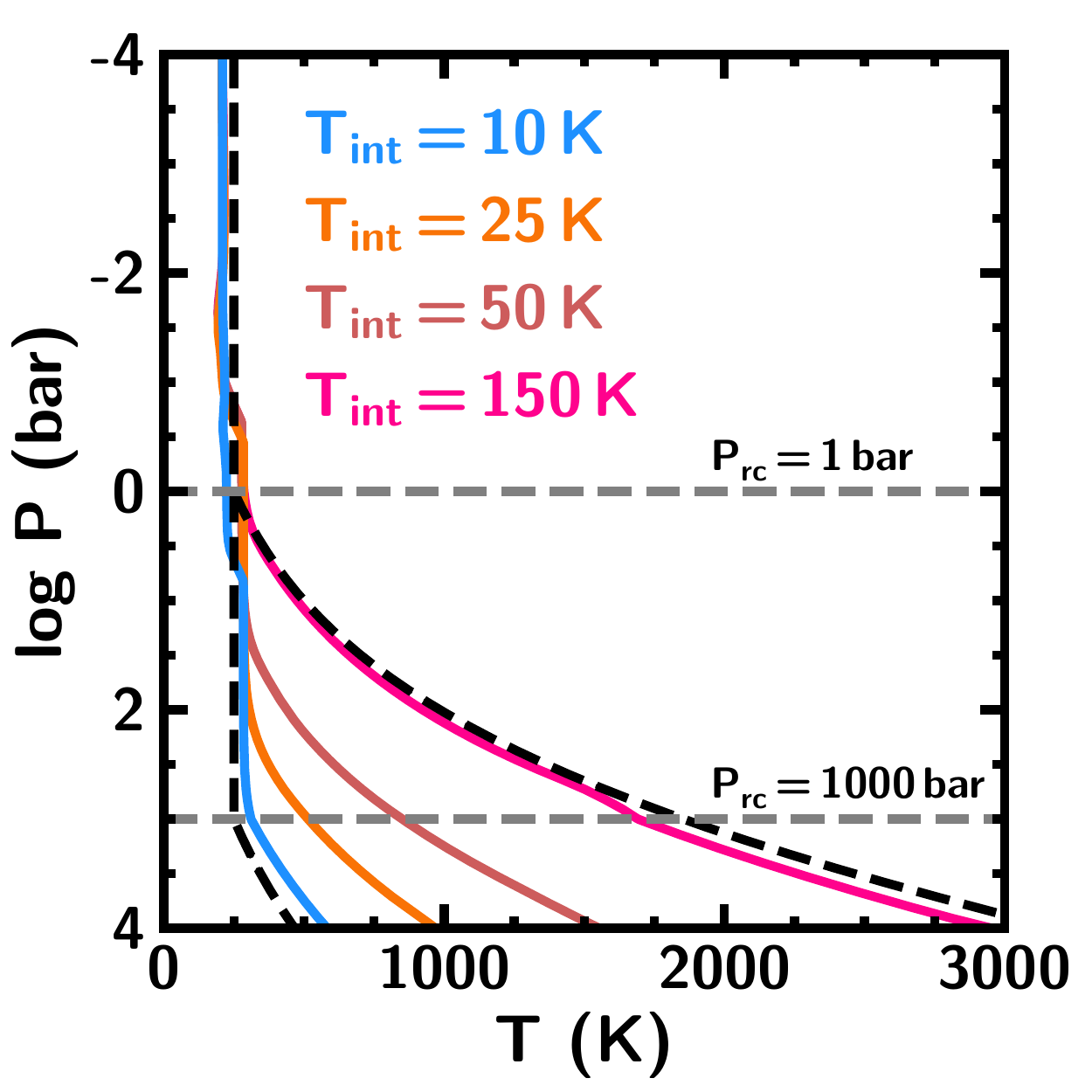}
    \caption{Comparison of analytic and self-consistent $P$--$T$ profiles. The solid curves show profiles from self-consistent models  \citep{Piette2020}, color-coded by the intrinsic temperature $T_{\rm int}$ used for the model. The self-consistent models span a broad range of possible internal energies for sub-Neptunes, assuming nominal planet properties based on the mini-Neptune K2-18b. Black dashed lines show isothermal/adiabatic profiles with $P_{\rm rc}=1\,$bar and $P_{\rm rc}=1000\,$bar. We consider these as end-member scenarios when modelling planets with H/He envelopes.}
    \label{fig:k2_pt_check}
\end{figure}

The resulting temperature profiles are shown in Figure \ref{fig:k2_pt_check}, and can be closely matched by isothermal/adiabatic profiles with $P_{\rm rc}$ lying between $1\,$bar and $1000\,$bar. We therefore take 1--1000$\,$bar to be a reasonable range of values for $P_{\rm rc}$ when considering a general sub-Neptune atmosphere in the remainder of this paper.

To further illustrate that a model incorporating an isothermal/adiabatic temperature profile can yield very similar results to a model with a temperature profile generated by a self-consistent model, we consider one of the interior models for the planet K2-18b from \citet{Madhu2020}, which used $P$--$T$ profiles produced by \textsc{genesis}. We take case 2 from that paper, which has a composition of 45\% Earth-like nucleus, 54.97\% H$_2$O, and 0.03\% H/He. We fit the temperature profile used for that case with an isothermal/adiabatic profile, finding best-fit parameters $T_0 = 300\,$K, $P_0=0.05\,$bar, and $P_{\rm rc}=3\,$bar. The model using an isothermal/adiabatic profile gives a radius of $2.613\,R_{\oplus}$ at the mean observed planet mass \citep[$8.63\,M_{\oplus}$,][]{Cloutier2019} which, like the model from \citet{Madhu2020}, agrees with the observed planetary radius to well within the observational uncertainty \citep[$2.610 \pm 0.087\,R_{\oplus}$;][]{Benneke2019}.

%%%%%%%%%%%%%%%%%%%%%%%%%%%%%%%%%%%%%%%%%%%%%%%%%%%%%%%%%%%%%%%%%%%%%%%%%%

\section{Model Validation} \label{section:validation}

%%%%%%%%%%%%%%%%%%%%%%%%%%%%%%%%%%%%%%%%%%%%%%%%%%%%%%%%%%%%%%%%%%%%%%%%%%

In this section we validate our model by reproducing results from a number of previous works concerning the internal structures of sub-Neptunes. We also examine the effect of a temperature-dependent interior on the radii of H$_2$O-rich planets and compare our results to other studies that have followed similar approaches.

\subsection{Comparison with previous studies}

\begin{figure}
\includegraphics[width=\columnwidth]{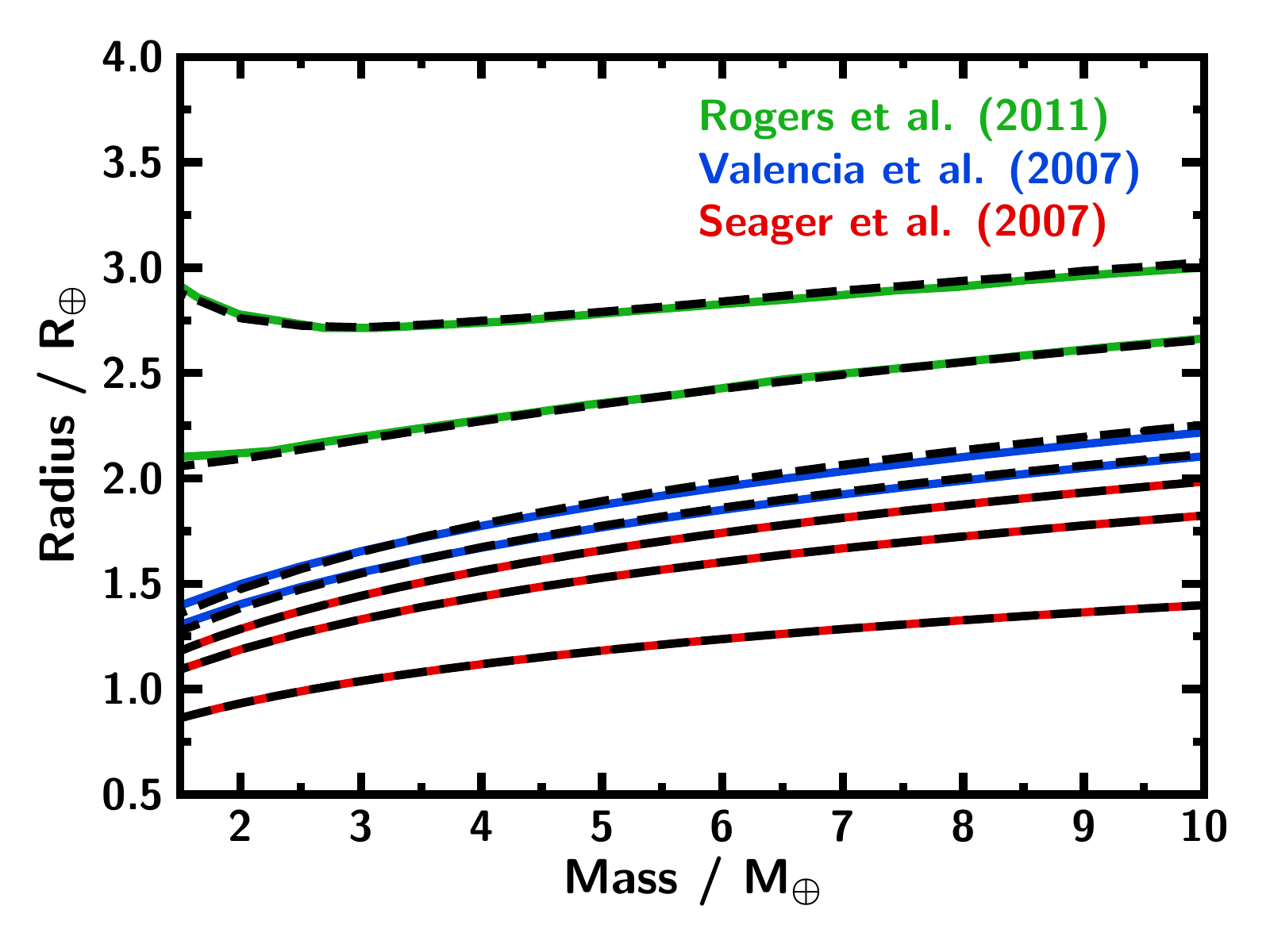}
    \caption{Comparison between $M$--$R$ relations produced by our model and previously published results. The red curves are taken from \citet{Seager2007} for planets of pure Fe, Earth-like and pure silicate composition. The blue curves are taken from \citet{Valencia2007} for planets with an Earth-like core and adiabatic H$_2$O layers of 30\% and 50\%. The green curves are taken from \citet{Rogers2011} for planets with a 33\% Earth-like nucleus, a $\sim$67\% isothermal H$_2$O layer and H/He envelopes of 0.1\% and 1\%. The black dashed lines show our model reproductions of each of these cases. All of our results agree with the previously published $M$--$R$ relations to within 0.05$R_{\oplus}$.}
    \label{fig:validation}
\end{figure}

Figure \ref{fig:validation} shows several $M$--$R$ relations from previous works alongside the results of our model given the same input parameters. We begin by reproducing the results of \citet{Seager2007} for isothermal planets comprised of iron and silicates. Since we use the same EOS for Fe and MgSiO$_3$ as that paper, we expect to find total agreement between their results and our model when considering planets consisting solely of iron and rock. We reproduce $M$--$R$ relations for three compositions shown in figure 4 of \citet{Seager2007}: a pure iron planet, a pure silicate planet, and a planet consisting of 32.5\% iron and 67.5\% silicates. In each of these cases our results agree completely with the published $M$--$R$ relations.

Next we consider $M$--$R$ relations shown in figure 3 of \citet{Valencia2007} for planets at $T_0 = 550$K with H$_2$O mass fractions of 10\% and 30\%. The H$_2$O layer consists of ice VII and X and follows an adiabatic temperature profile. The inner layers of the planet are made up of silicates and iron in a 2:1 ratio. \citet{Valencia2007} do not provide a surface pressure for their models, so in order to reproduce their results we take $P_0 = 10^{10}\,$Pa, since this forces the phase at the surface to be ice VII rather than liquid. The ice VII EOS used in our models differs slightly from \citet{Valencia2007}, however we find good agreement between the two sets of $M$--$R$ curves. Across all masses and compositions considered, the largest discrepancies in radius are below $0.04R_{\oplus}$, smaller than the observational uncertainties of even the best super-Earth radius measurements ($\sim$0.1$R_{\oplus}$). The small differences that do appear are likely due to the different formulation of the ice VII EOS and the fact that \citet{Valencia2007} incorporate conduction in the mantle.

Finally we compare our model to equilibrium models of planets with H/He envelopes from \citet{Rogers2011}. We consider $M$--$R$ relations shown in figure 4 of that paper for planets with $T_{eq}=500$K and H/He mass fractions of 0.1\% and 1\%. The underlying composition in each case is 10\% Fe, 23\% silicates and 67\% H$_2$O. \citet{Rogers2011} did not use a temperature-dependent H$_2$O EOS, instead adopting the isothermal EOS from \citet{Seager2007}, and so for our reproduction we take an isothermal temperature profile at 300$\,$K in the H$_2$O layer. \citet{Rogers2011} also used the H/He EOS from \citet{Saumon1995}, which differs from the \citet{Chabrier2019} EOS at high densities ($\rho > 50\,$kg$\,$m$^{-3}$). The temperature profile used in the H/He envelope is analytic, taking the form of Equation \ref{eq:Guillot} with $\gamma = 0.6 \sqrt{T_{\rm irr}/2000{\rm K}}$, $f=1/4$ and $T_{\rm int} = (L_p/4\pi R_p^2 \sigma)^{1/4}$, where $L_p$ is the intrinsic luminosity of the planet and $\sigma$ is the Stefan-Boltzmann constant. For the models reproduced here, $L_p$ is determined by fixing $L_p/M_p = 10^{-10.5}$ W kg$^{-1}$. For the purposes of this reproduction we do not correct for the transit radius effect. Again we find good agreement between our models and those of \citet{Rogers2011}, with maximum discrepancies less than $ 0.05R_{\oplus}$. These may be a result of the differing H/He EOS. Regardless, our $M$--$R$ relations and all those from previous studies shown here agree to well within typical observational uncertainties for super-Earths.

\subsection{Mass--radius relations} \label{subsection:mr}

Here we use our model to produce $M$--$R$ relations for sub-Neptune exoplanets with varying temperature structures and compositions, and compare these to previous work in the field.

\subsubsection{Water worlds with no H/He envelope}
\label{subsection:h2o_mr}

\begin{figure*}
\includegraphics[width=\textwidth]{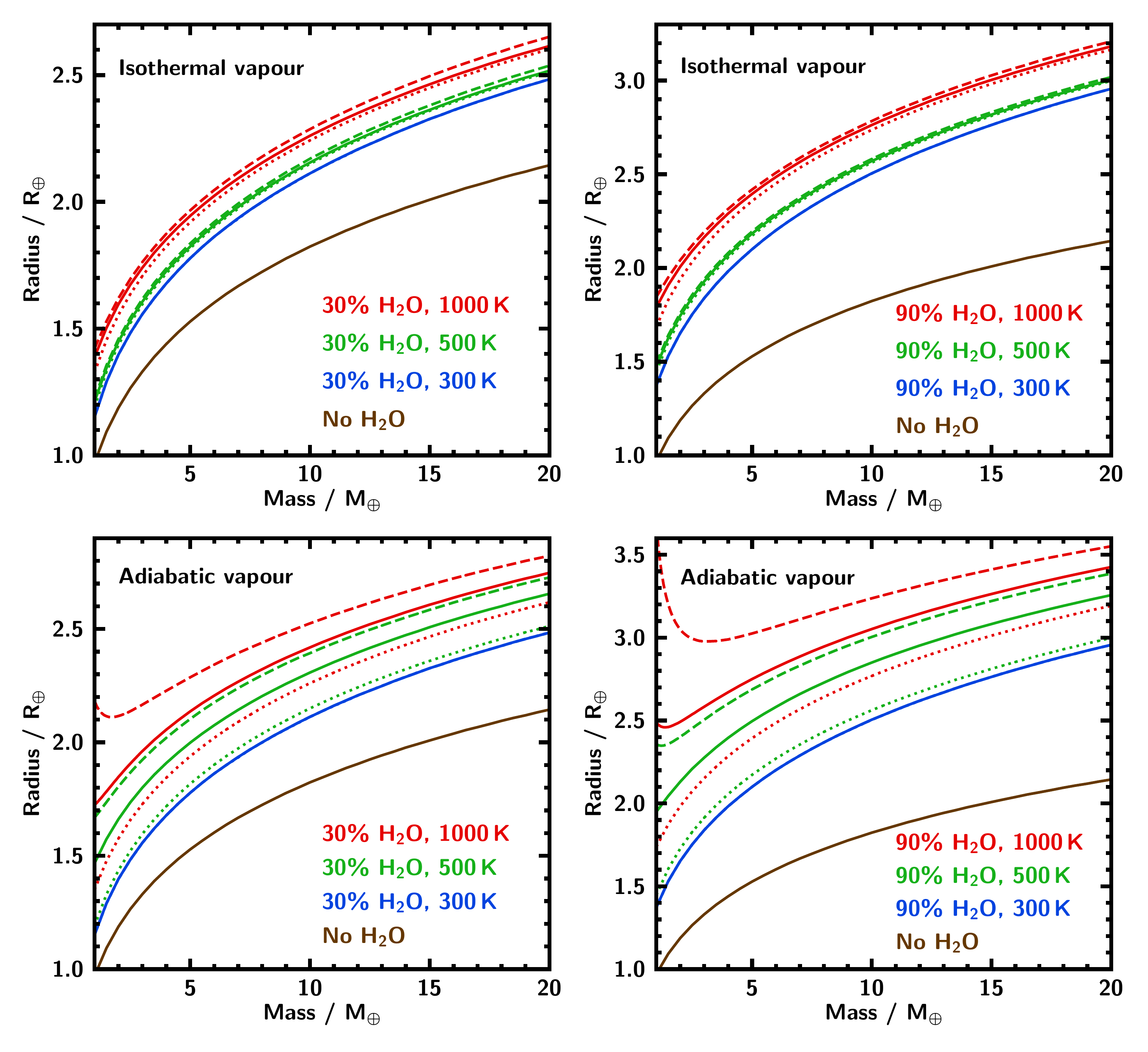}
    \caption{$M$--$R$ relations for water worlds without H/He envelopes. The model planets consist of a H$_2$O layer of 30\% (left) or 90\% (right) above an Earth-like nucleus (1/3 iron, 2/3 silicates by mass). All planets follow an adiabatic temperature profile in the liquid, supercritical and ice phases of the H$_2$O layer. In the top panel, planets with water vapour follow an isothermal temperature profile in the vapour layer, whereas in the bottom panel the temperature profile in this layer is also adiabatic. The line colour denotes the photospheric temperature $T_0$ and the line style indicates the photospheric pressure: solid lines show models with $P_{0}=1\,$bar, dotted lines show models with $P_{0}=100\,$bar and dashed lines show models with $P_{0}=0.1\,$bar. Increasing $T_0$ from 300--1000$\,$K increases the planet radius considerably, with a more pronounced effect if the vapour layer is adiabatic. For planets with isothermal vapour layers, changing $P_0$ does not strongly affect the radius, but decreasing $P_0$ for a planet with an adiabatic vapour layer can substantially increase its radius.}
    \label{fig:h2o_summary}
\end{figure*}

Figure \ref{fig:h2o_summary} shows $M$--$R$ curves for H$_2$O-rich planets with no H/He envelope. We assume that the underlying nucleus is  Earth-like in composition (1/3 Fe, 2/3 MgSiO$_3$ by mass). For the liquid, ice and supercritical phases of H$_2$O we assume an adiabatic temperature profile. We consider two end-member temperature profiles in the vapour phase: an isothermal profile and an adiabatic profile. We consider the effect of several different parameters on the planetary radius, including the temperature $T_0$ and pressure $P_0$ at the photosphere as well as the water mass fraction $x_{\rm H_2O}$. The photospheric pressure and temperature of a planet can be constrained from observations: $T_0$ is closely related to the planetary effective temperature, which can be estimated from the orbital separation of the planet and the luminosity of the host star. $P_0$ is defined as the pressure at the photosphere (where $R=R_p$). It can be retrieved from an atmospheric spectrum \citep{WelbanksMadhu2019,Nixon2020} and subsequently used as a boundary condition when analysing a planet's internal structure \citep{Madhu2020}.

Our findings agree with previous studies of water worlds. Similarly to \citet{Thomas2016} we find that variations in $T_0$ can result in significant changes to planetary radii, which are even more drastic if the water vapour layer is assumed to be adiabatic. We also find an upper limit of $\sim$3$R_{\oplus}$ for planets with a solid or liquid surface, in agreement with \citet{Zeng2019}. The effects of changing various parameters are discussed in greater detail in Appendix \ref{appendix:a}.

\subsubsection{Water worlds with H/He envelopes} \label{subsec:hhe_mr}

\begin{figure*}
\includegraphics[width=\textwidth]{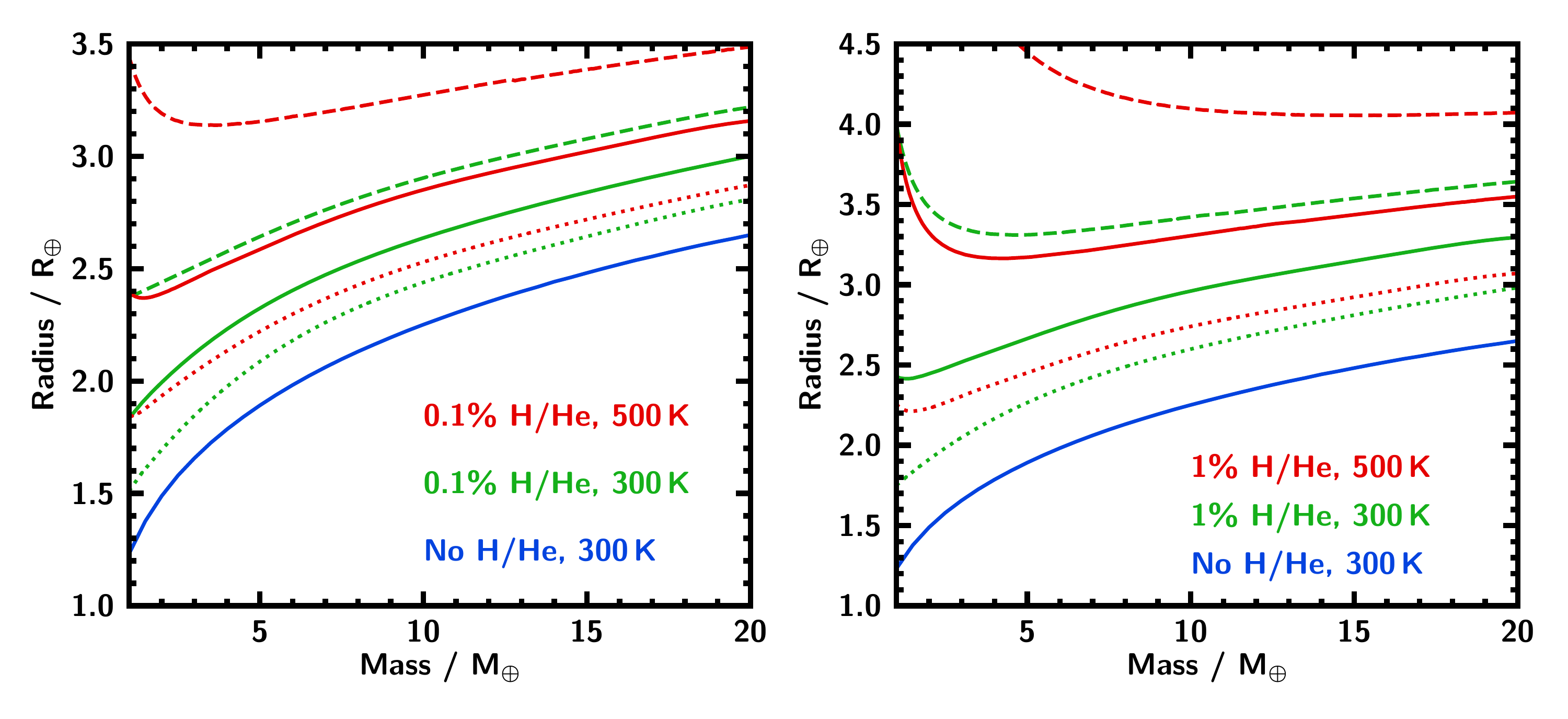}
    \caption{$M$--$R$ relations for planets with H/He envelopes. The line style indicates the location of the radiative-convective boundary: solid lines show models with $P_{\rm rc}=10\,$bar, dotted lines show models with $P_{\rm rc}=1000\,$bar and dashed lines show models with $P_{\rm rc}=1\,$bar. Even a small amount of H/He can inflate the radius of a planet substantially, and this effect is more pronounced at higher temperatures. Increasing $P_{\rm rc}$ decreases the planet radius.}
    \label{fig:hhe_summary}
\end{figure*}

Next we model planets with an Earth-like nucleus (1/3 iron, 2/3 silicates) and an adiabatic H$_2$O layer of equal mass to the nucleus underneath a H/He envelope. We adopt isothermal/adiabatic temperature profiles for the H/He layer as described in Section \ref{subsubsec:ia}, meaning the temperature profile is determined by three parameters: $T_0, P_0$ and $P_{\rm rc}$. 

Figure \ref{fig:hhe_summary} shows $M$--$R$ relations for planets with different photospheric temperatures, radiative-convective boundaries and H/He mass fractions. The main factors governing the $M$--$R$ relation for water worlds with H/He envelopes are the temperature profile in the envelope and the mass fraction of H/He. Many features of the $M$--$R$ relations presented in this figure are well-documented in the literature \citep[e.g.][]{Rogers2011,Lopez2014}, such as the inflation of radii at low masses, which is a result of low surface gravity increasing the atmospheric scale height, and the significant effect of $T_0$ on the planetary radius. Further discussion of the effects of changing the temperature profile and H/He mass fraction of the planet can be found in Appendix \ref{appendix:a}.

%%%%%%%%%%%%%%%%%%%%%%%%%%%%%%%%%%%%%%%%%%%%%%%%%%%%%%%%%%%%%%%%%%%%%%%%%%

\section{Results} \label{section:results}

%%%%%%%%%%%%%%%%%%%%%%%%%%%%%%%%%%%%%%%%%%%%%%%%%%%%%%%%%%%%%%%%%%%%%%%%%%

Here we present our results exploring in detail the internal phase structures of H$_2$O-rich super-Earths and mini-Neptunes. We investigate four key aspects. We begin by calculating the range of possible ocean depths on planets with a large H$_2$O component across a wide range of possible bulk properties and temperature structures. We allow for the full extent of the liquid phase of H$_2$O, reaching temperatures as high as 647 K at pressures from 218--7$\times$10$^4\,$bar. Next we determine the range of masses, radii and surface conditions for which a mini-Neptune with a H/He envelope may possess a liquid water ocean underneath. We also consider how different temperature profiles can affect the phase structures of water worlds, investigating planets with ice and vapour surfaces as well as those with surface oceans. Finally, we consider how miscibility of H/He and H$_2$O within a planet's envelope might affect the $M$--$R$ relation.

\subsection{Depth of oceans on water worlds} \label{subsection:ocean}

\begin{figure*}
\includegraphics[width=\textwidth]{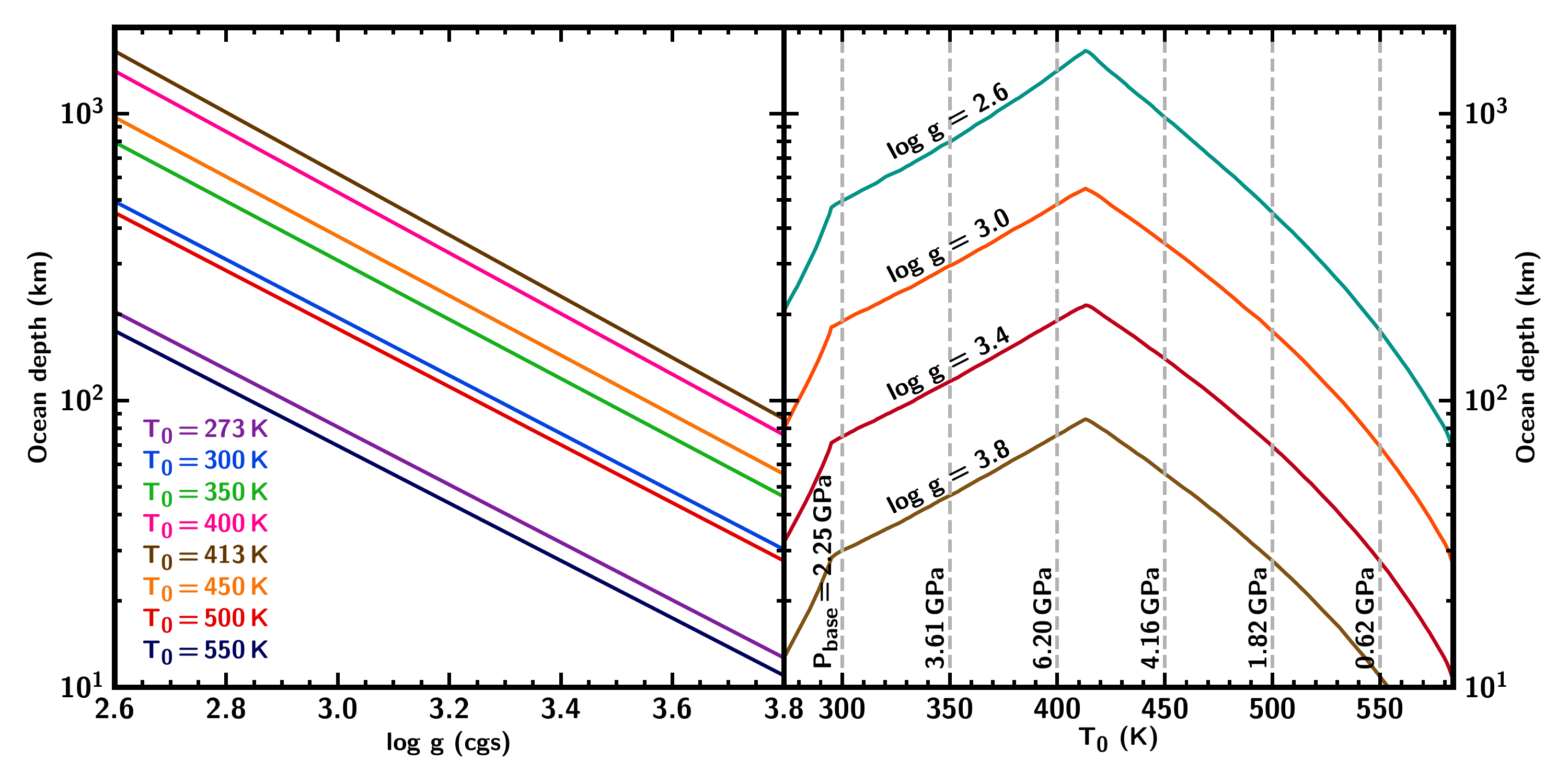}
    \caption{Extent of oceans on H$_2$O-rich planets with different surface gravities $g$ and temperatures $T_0$. The left-hand panel shows the ocean depth against $\log g$ at different values of $T_0$ from 273--550$\,$K. Ocean depth is inversely proportional to surface gravity for planets without gaseous envelopes. The right-hand panel shows ocean depth against $T_0$ for different values of $\log g$ (cgs) from 2.6--3.8. The grey dashed lines indicate the pressure at the bottom of the ocean, $P_{\rm base}$, which depends on $T_0$. A higher $P_{\rm base}$ leads to a deeper ocean.}
    \label{fig:ocean_depth_g_t0}
\end{figure*}

Our primary goal is to estimate the possible depths of liquid H$_2$O layers in water-dominated planets over a range of surface conditions. For reference, the Earth's oceans have an average depth of 3.7$\,$km, extending to 11$\,$km at their deepest point \citep{Charette2010,Gardner2014}. Given that the Earth's H$_2$O mass fraction is $\sim$0.02$\%$, one might expect that planets with a significant portion of their mass in H$_2$O should possess deeper oceans. However, it is also the case that much of the H$_2$O layer of a water world can be in supercritical or high-pressure ice phases due to high pressures in the interior. Therefore, the possible extent of an ocean is not simply limited by the amount of H$_2$O available.

Here we investigate the range of ocean depths that may be achieved in the case where a planet has enough H$_2$O that the size of the ocean is not limited by water mass fraction. While other works have explored this to some extent as discussed previously, for this study we consider planets across a wider range of masses, compositions and surface conditions. We aim to determine which parameters are important in controlling the extent of an ocean, and how the ocean depth varies across the full parameter space encompassed by super-Earths and mini-Neptunes that may host H$_2$O layers. We  parametrise the temperature profile of the H$_2$O layer by assuming an adiabatic profile with a H$_2$O surface of $P_0=100\,$bar, where $T_0=T(P_0)$ is a free parameter that determines the adiabat. While the thermodynamic conditions at the surface of the H$_2$O layer are determined by numerous factors, including the level of irradiation received from the host star, the planet's intrinsic temperature $T_{\rm int}$, and atmospheric properties such as opacity from molecular and atomic chemical species, by varying $T_0$ across the full set of possibilities for a liquid surface we can encapsulate all feasible cases.

Before exploring the extent of oceans found by our internal structure models, we can examine the approximate behaviour of ocean depths by returning to the equation of hydrostatic equilibrium (Equation \ref{eq:hydro_eqm}), but with radius rather than mass as the independent variable:

\begin{equation}
    \frac{dP}{dR} = -\rho g,
\end{equation}
where $g$ is the gravitational acceleration. Consider a planet with a liquid H$_2$O surface. From Section \ref{subsec:h2o_phase} we can see that the ocean is unlikely to constitute a large portion of the planet interior, and so we can assume that $g$ is constant throughout the ocean. If we also consider the density of liquid water to remain constant (see Figure \ref{fig:h2o_eos}), then we find that the depth of the ocean is proportional to change in pressure from the surface to the base. If the pressure at the base of the ocean is much larger than the surface pressure, then we expect the base pressure $P_{\rm base}$ to strongly affect the ocean depth. For a planet with a large H$_2$O mass fraction, the value of $P_{\rm base}$ is determined by the location where the adiabatic $P$--$T$ profile crosses from liquid to either supercritical water or high-pressure ice. This in turn depends on the chosen surface temperature. Therefore we would expect the key parameters in determining the ocean depth to be the gravity and temperature at the ocean's surface.

We now turn to the full models to explore this in more detail. Our analysis above suggests that planets with the same surface temperature and gravity should have approximately the same ocean depth, regardless of other factors like the planet mass, radius and composition. We find this to be the case when using the full models. For example, consider two ocean planets corresponding to points from the $M$--$R$ curves from Figure \ref{fig:h2o_summary}: a 5$M_{\oplus}$ planet with 30\% H$_2$O by mass, which has a radius of 1.78$R_{\oplus}$, and a 10$M_{\oplus}$ planet with 90\% H$_2$O by mass, whose radius is 2.51$R_{\oplus}$. Both planets have a surface temperature of 300$\,$K and the same surface gravities, $\log g \, \rm{(cgs)}=3.19$, despite their differing masses, radii and compositions. We find that the oceans on both planets are 125$\,$km deep. Similarly, we find that ocean depth is not strongly affected by the surface pressure or the presence of an atmosphere. However, we note for a planet with a gaseous envelope, the ocean depth is determined by the gravity at the surface of the ocean rather than the gravity with the envelope included.

Having confirmed that surface temperature and gravity are the main parameters which determine the extent of an ocean, we explore the ($g,T_0$) parameter space in order to find the range of possible ocean depths for H$_2$O-rich planets. We consider values of log $g$ (cgs) ranging from 2.6--3.8. A 90\% H$_2$O planet at 1$M_{\oplus}$ with $T_0=500\,$K, has $\log g = 2.67$, which gives a reasonable lower bound for the surface gravity of a water world. The upper limit of $\log g = 3.8$ corresponds to an extreme case of a 20$M_{\oplus}$ iron planet with a thin H$_2$O layer. For reference, the water-rich planets considered later in Figure \ref{fig:h2o_cases} all have $\log g \approx 3.2$. We also consider surface temperatures from 273--584$\,$K, spanning the liquid phase assuming a surface pressure of 100$\,$bar.

Figure \ref{fig:ocean_depth_g_t0} shows the depths of oceans for planets across this parameter space. As expected from the arguments at the beginning of this section, we find that ocean depth is inversely proportional to gravity and approximately directly proportional to $P_{\rm base}$, which is determined by the surface temperature. From the adiabatic temperature profiles shown in Figure \ref{fig:water_phases} we would expect that the optimum surface temperature in order to maximise $P_{\rm base}$, and hence ocean depth, lies between 350 and 450$\,$K. From the grid of models we see that the surface temperature which maximises ocean depth assuming an adiabatic temperature profile starting from 100$\,$bar is 413$\,$K. At this temperature, and a minimal $\log g =2.6$, the ocean depth is 1654$\,$km, about 450 times the average depth of the Earth's ocean \citep[3.7$\,$km,][]{Charette2010}. This can be considered an extreme upper limit to the depth of an ocean on a water world. 

In scenarios that have been explored previously, our results are in agreement with past work. For example, for a 6$M_{\oplus}$ planet with 50\% H$_2$O we find ocean depths of 66$\,$km for $T_0=280\,$K and 125$\,$km for $T_0=303\,$K, which are similar to the values of 72$\,$km and 133$\,$km from \citet{Leger2004}.

Looking at the broader parameter space, the models show that a wide variety of ocean depths are possible. For example, a water world with a 300$\,$K surface can have an ocean depth from 30--500$\,$km, or 8--135 times deeper than the Earth's ocean, depending on its mass and composition. For a given planet mass, higher H$_2$O mass fractions lead to deeper oceans, since planets with more H$_2$O relative to iron and silicates will have lower surface gravities allowing for a more extended liquid water layer. A $1M_{\oplus}$ planet with a 30\% water layer and $T_0=300\,$K has log $g=2.83$ and an ocean depth $R_{\rm ocean}=283\,$km, about 76 times deeper than the average depth on Earth, whereas a $1M_{\oplus}$ planet with the same $T_0$ but a water mass fraction of 90\% has log $g=2.70$ and $R_{\rm ocean}=388\,$km. We can also see that for a fixed composition, more massive planets have higher surface gravities and so $R_{\rm ocean}$ decreases as the mass of the planet increases. For example, a 20$M_{\oplus}$ planet with a 30\% water layer and $T_0=300\,$K has log $g=3.50$ and $R_{\rm ocean}=59\,$km.

\subsection{Potential for liquid water on mini-Neptunes}
\label{subsec:hhe_phase}

We now consider the extent to which planets possessing H/He envelopes may host liquid H$_2$O at their surface. Assuming that the H/He and H$_2$O layers of the planet do not mix, the pressure and temperature at the H$_2$O-H/He boundary (HHB) depends on the H/He mass fraction and the atmospheric temperature profile \citep{Madhu2020}. This in turn determines the phase of H$_2$O at the HHB. The liquid phase of H$_2$O is accessible at temperatures up to 647$\,$K and pressures up to $7.3 \times 10^4\,$bar (see Figure \ref{fig:h2o_eos}).

\begin{figure}
\includegraphics[width=\columnwidth]{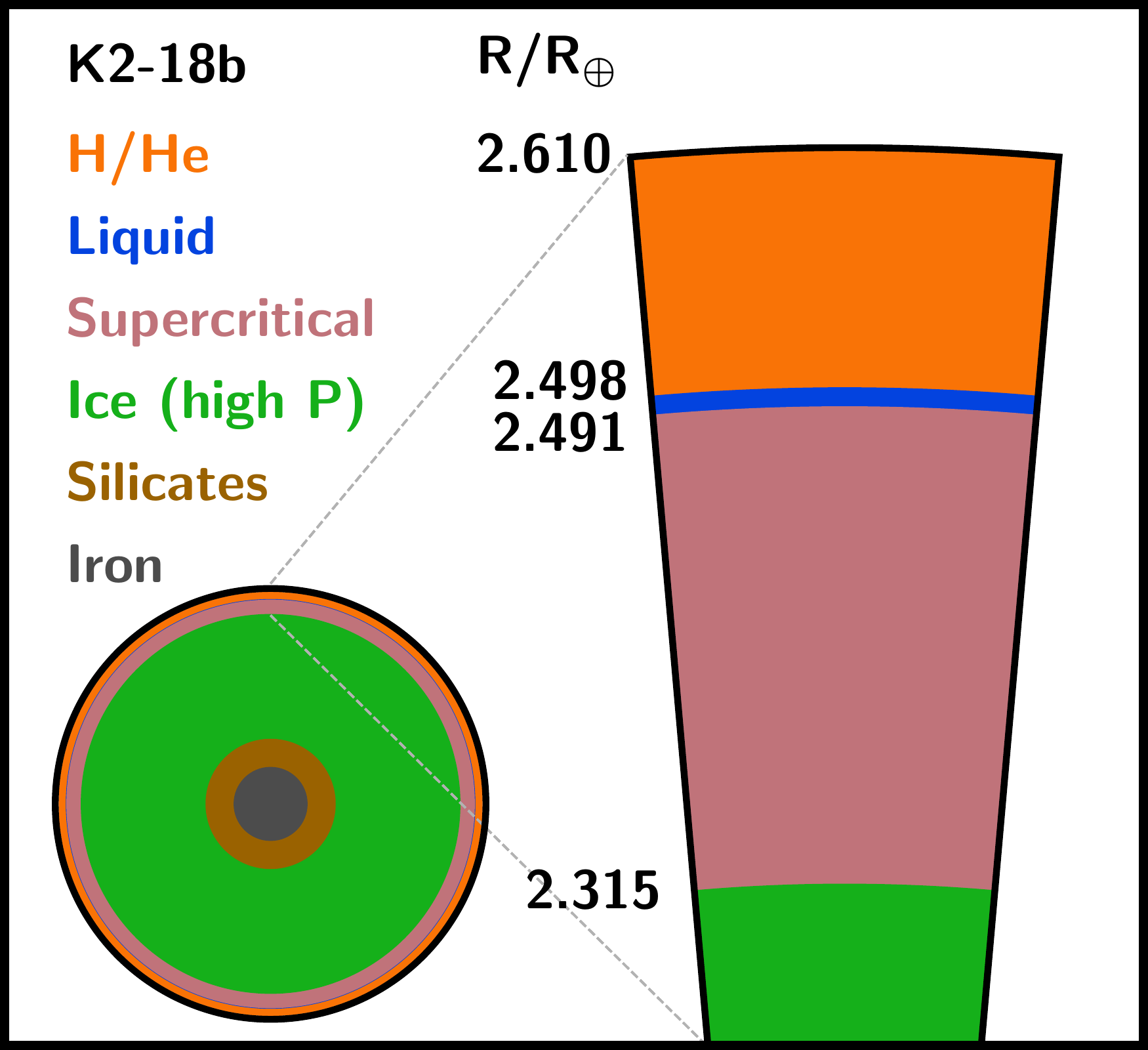}
    \caption{Internal structure for one possible interior composition of K2-18b, corresponding to Case 3 from \citet{Madhu2020} with a H/He mass fraction of $6 \times 10^{-5}$, an Earth-like nucleus of 10\% by mass and a $\sim$90\% H$_2$O layer. A layer of liquid water is present directly beneath the H/He envelope.}
    \label{fig:k218b_case3}
\end{figure}

\citet{Madhu2020} demonstrated that the mini-Neptune K2-18b, despite having a mass and radius indicative of a planet with a substantial gaseous envelope, could have liquid water at the HHB. This scenario arises in the case where K2-18b consists mostly of H$_2$O, with a small Earth-like nucleus and a low but non-negligible H/He mass fraction. Figure \ref{fig:k218b_case3} shows the internal structure of one such model planet, the equivalent of Case 3 from \citet{Madhu2020}, which has a nucleus mass fraction of 10\% and a H/He mass fraction of $6 \times 10^{-5}$, with the rest of the planet consisting of H$_2$O. The phase structure of the H$_2$O layer is also shown.

These results hint at a region in $M$--$R$ space where planets must host non-negligible gaseous envelopes to explain their inflated radii, but could still possess a liquid H$_2$O layer. Here we aim to determine the extent of this region of the parameter space and investigate which, if any, other observed planets fall into this category. To begin, we construct the $M$--$R$ curve for a theoretical planet with an Earth-like nucleus of 10\% and a H$_2$O layer of 90\% by mass, with a surface temperature of 300$\,$K. We consider this to be an upper limit to the radius of a planet with surface liquid water and no H/He envelope. We choose this composition since a 100\% H$_2$O planet is unrealistic from a planet formation perspective (see Section \ref{subsection:h2o_mr}). Next we give the planet a H/He envelope, which we assume is fully differentiated from the H$_2$O layer (see Section \ref{subsection:mix} for a discussion of mixed envelopes). The pressure and temperature at the HHB increases with the mass of the envelope, since we assume a temperature profile consisting of an isotherm and an adiabat; i.e., the temperature either remains constant or increases monotonically with depth (see Section \ref{subsec:hhe_mr}). Therefore, for a habitable-zone temperature planet with a fixed core mass fraction, there must be a maximum amount of H/He that it can possess before the HHB becomes too hot for liquid water. For a given mass and temperature profile we can solve for this H/He mass fraction and therefore obtain the maximum radius of a planet that could host liquid water.

We find solutions for a finely-spaced grid of masses and a range of temperature profiles. For illustration, we consider planets from 1--20$M_{\oplus}$ and nominally assume a photospheric pressure $P_0=0.1\,$bar, with photospheric temperatures $T_0$ ranging from 100--647$\,$K and $P_{\rm rc}$ from 1--1000$\,$bar; the parameter ranges are motivated by the $P$--$T$ profiles shown for a wide range of mini-Neptune atmospheres in \citet{Piette2020}, as discussed in Section \ref{subsec:choice} and Figure \ref{fig:k2_pt_check}. Although we consider photospheric temperatures up to the critical temperature of H$_2$O, i.e. the maximum temperature at which liquid H$_2$O can exist, we note that for planets with $T_0 \geq 319\,$K at our chosen $P_0$ an inflated radius could also be attributed to a steam atmosphere (see Section \ref{subsec:h2o_phase}). 

\begin{figure*}
\includegraphics[width=\textwidth]{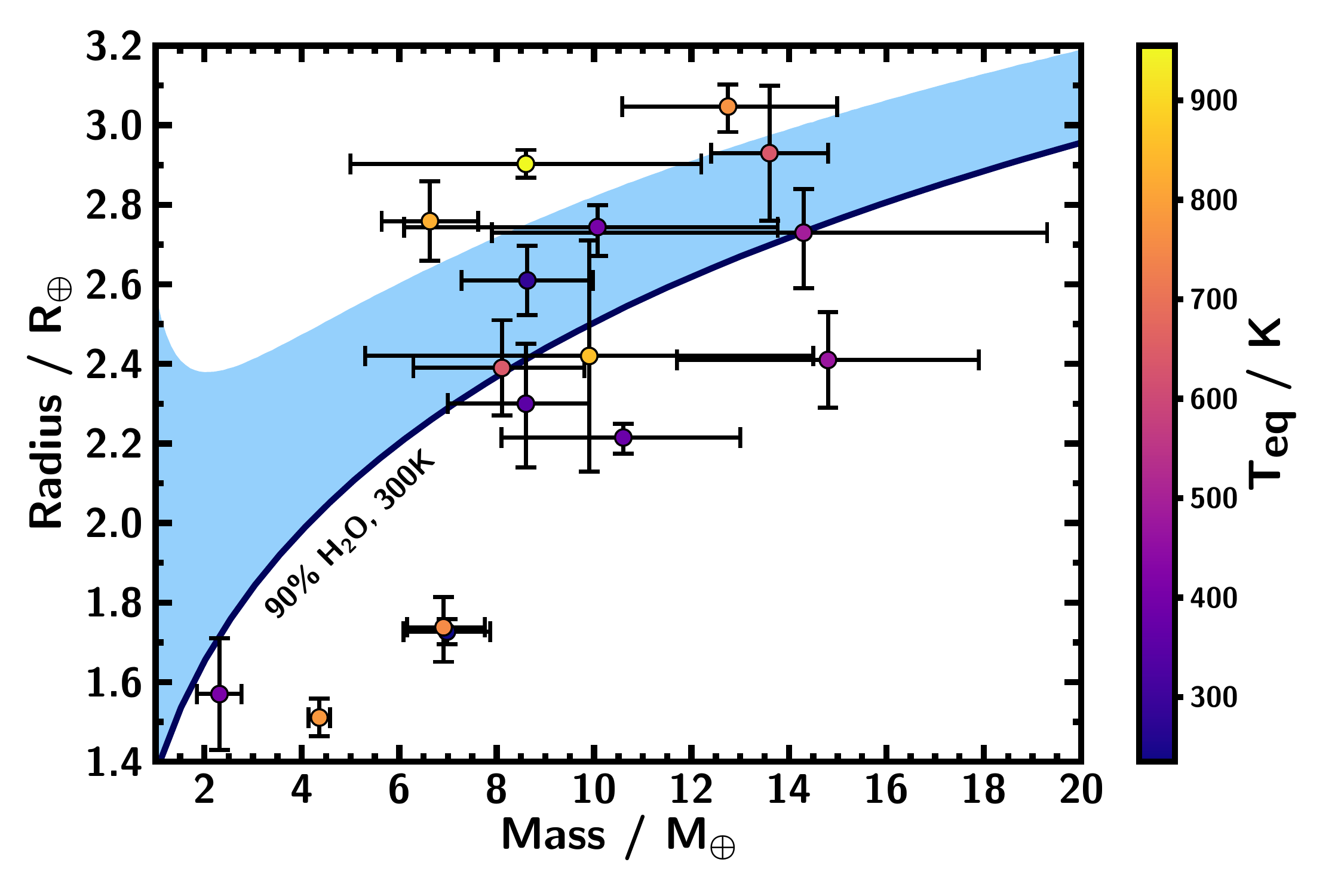}
    \caption{$M$--$R$ diagram showing the range of planet masses and radii consistent with a H/He envelope and liquid H$_2$O at the HHB (light blue shaded region). Temperatures at the HHB can lie anywhere within the liquid phase of water, which extends to 647$\,$K at high pressures. The dark blue line shows a theoretical $M$--$R$ relation for a planet consisting of a 10\% Earth-like nucleus and a 90\% H$_2$O layer with a 300$\,$K surface temperature, which we take as the upper limit for a planet with no H/He envelope. We also show planets whose masses and radii have been reported with at least 2$\sigma$ confidence, with $T_{\rm eq} \leq 1000\,$K. $T_{\rm eq}$ is indicated by the colour of each planet. Data on planetary masses, radii and equilibrium temperatures is taken from the NASA Exoplanet Archive.}
    \label{fig:liquid_mr}
\end{figure*}

Our results are shown in Figure \ref{fig:liquid_mr}, which shows the region of $M$--$R$ space where a planet hosting a H/He envelope could possess a liquid water ocean. We also show measured masses and radii of planets near this region. We find that for a given $T_0$ and $P_{\rm rc}$, there is a maximum total mass of H/He envelope, $M_{\rm env}$, that can allow for liquid water at the HHB. This maximum $M_{\rm env}$ increases as the HHB becomes deeper, meaning the shaded region in Figure \ref{fig:liquid_mr} represents planets with a maximal $P_{\rm rc}=1000\,$bar. For planets whose atmospheres have a radiative-convective boundary at a lower pressure, the temperature starts to increase from higher in the atmosphere. Therefore, the envelope must be smaller to maintain liquid water at the HHB. For example, consider a planet with $T_0=300\,$K. If $P_{\rm rc}=1000\,$bar, then a $10M_{\oplus}$ planet can host up to $7.0 \times 10^{-2}M_{\oplus}$ of H/He while retaining liquid water at the HHB. However, if $P_{\rm rc}=10\,$bar, then the maximal $M_{\rm env} = 4.1 \times 10^{-4}M_{\oplus}$. The mass of envelope permitted also decreases as the photospheric temperature increases, following an approximate power law behaviour: for example, for $M_p=10M_{\oplus}$ and $P_{\rm rc}=10\,$bar we have $(M_{\rm env} / M_{\oplus}) \approx 3.3 \times 10^6\, (T_{0}/K)^{-4}$. $M_{\rm env}$ only weakly depends on the total planet mass $M_p$: for example, at $T_0=300\,$K, $P_{\rm rc} = 10\,$bar, a $20M_{\oplus}$ planet has at most $M_{\rm env} = 3.4 \times 10^{-4}M_{\oplus}$ while a $1M_{\oplus}$ planet has up to $M_{\rm env} = 5.8 \times 10^{-4}M_{\oplus}$ while retaining a surface ocean. This behaviour also follows an approximate power law, with $M_{\rm env} \approx 6 \times 10^{-4}\, M_p^{-0.18}$ in this case. Despite the maximum permissible mass of the H/He envelope decreasing with increasing $T_0$, the increase in radius is approximately the same across all values of $T_0$ considered. Although a hotter H/He envelope is less dense and therefore should be more inflated, the maximum $P_{\rm HHB}$ that can permit a water ocean also decreases as $T_0$ increases, cancelling out the increase in radius from the warmer, less dense envelope. The maximum permitted H/He envelope across all temperature profiles considered increases the radius of the planet by 0.23--1.19$R_{\oplus}$, with larger envelopes seen for lower mass planets due to the smaller surface gravity.

Our results hold for a H$_2$O mass fraction of 0.9 and can be treated as an upper limit for the radius of a planet with a liquid water ocean. For planets with less H$_2$O relative to the nucleus, the increase in radius permitted from a H/He envelope while retaining a liquid HHB is smaller, since the planet has a higher surface gravity. For example, in the 90\% H$_2$O case, a 10$M_{\oplus}$ planet has $R_p=2.51R_{\oplus}$ and therefore $\log g = 3.192$ (cgs). For $P_{\rm rc}=10\,$bar, the maximum increase in radius allowed from a H/He envelope while retaining an ocean is 0.238$R_{\oplus}$. A planet with the same mass but with only 50\% H$_2$O (and $\sim$50\% nucleus) has $\log g = 3.287$, and the maximum increase in radius for this planet is 0.156$R_{\oplus}$. 

From Figure \ref{fig:liquid_mr} we see that there are several planets other than K2-18b whose observed masses and radii could be explained with a small H/He envelope, possibly with liquid H$_2$O at the HHB. However, most of these planets have equilibrium temperatures high enough that their radii could be also be explained with a steam atmosphere. One planet which falls into this category is Kepler 20-d \citep{Gautier2012}, which has a mass of $10.07^{+3.97}_{-3.70}M_{\oplus}$ and a radius of $2.744^{+0.073}_{-0.055}R_{\oplus}$, with an equilibrium temperature of $401 \pm 13\,$K \citep{Buchhave2016}. A primarily H$_2$O planet with a steam atmosphere would be consistent with these measurements, as would a planet with a H/He envelope of up to a few per cent. Due to its surface temperature, whether such a planet could host a liquid H$_2$O layer is very sensitive to the choice of temperature profile: while there are solutions allowing for liquid water when using the temperature profiles considered here, the atmosphere of Kepler 20-d has not been observed, and so it is not known whether the hydrogen-rich temperature profiles used in this section would be appropriate to model this particular planet. This problem will be alleviated as more mini-Neptune atmospheres are observed in the future.

\subsection{Diversity of water world phase structures} \label{subsec:h2o_phase}

Our results thus far have focused on planets with a liquid water component, however in reality these planets represent just a fraction of the possible phase structures of water worlds. Here we discuss the wide variety of H$_2$O phase structures that may be present on water-rich planets. To achieve this we examine each phase of H$_2$O that can be found at the planet's surface and determine the various phase structures that the H$_2$O layer may possess depending on its internal temperature profile. 

\begin{figure}
\includegraphics[width=\columnwidth]{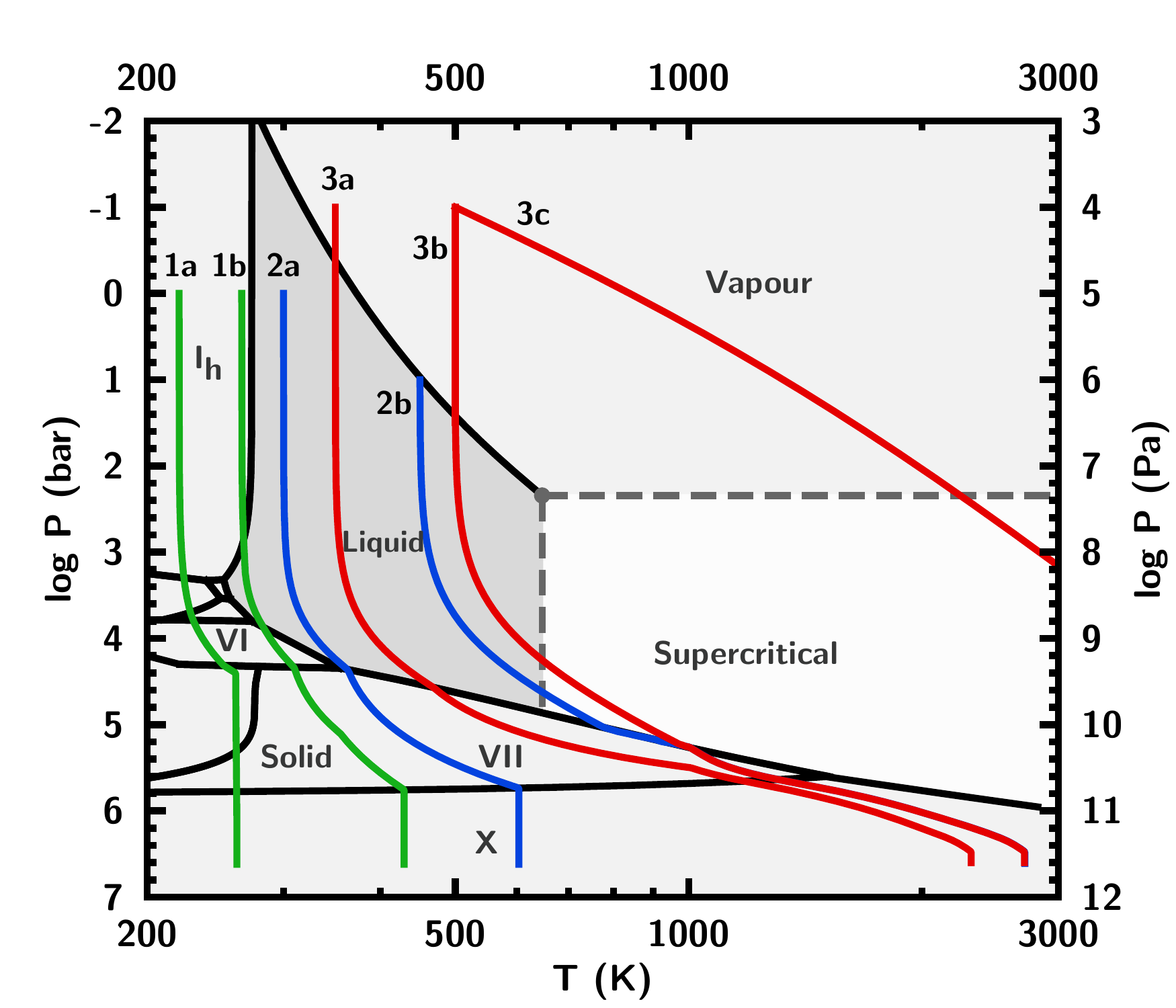}
    \caption{$P$--$T$ profiles of the planets shown in Figure \ref{fig:h2o_cases}. An adiabatic profile is used for all phases other than vapour. In the vapour phase, an isothermal profile is used in cases 3a and 3b, and an adiabatic profile is used in case 3c.}
    \label{fig:water_phases}
\end{figure}

The phase structure of the H$_2$O layer of a planet depends strongly on the choice of internal temperature profile. Most of the planet's interior will be convective and so we use an adiabatic temperature profile for the liquid, ice and supercritical phases \citep{Valencia2007,Thomas2016_thesis}. We consider two end-member cases for vapour layers: an isothermal or an adiabatic profile. Since the true temperature profile of a steam atmosphere is likely to lie between these extremes, this should allow us to explore the set of possible phase structures for water-rich planets with steam atmospheres. 

\begin{figure*}
\centering
\begin{center}$
\begin{array}{ccc}
\includegraphics[width=0.325\textwidth]{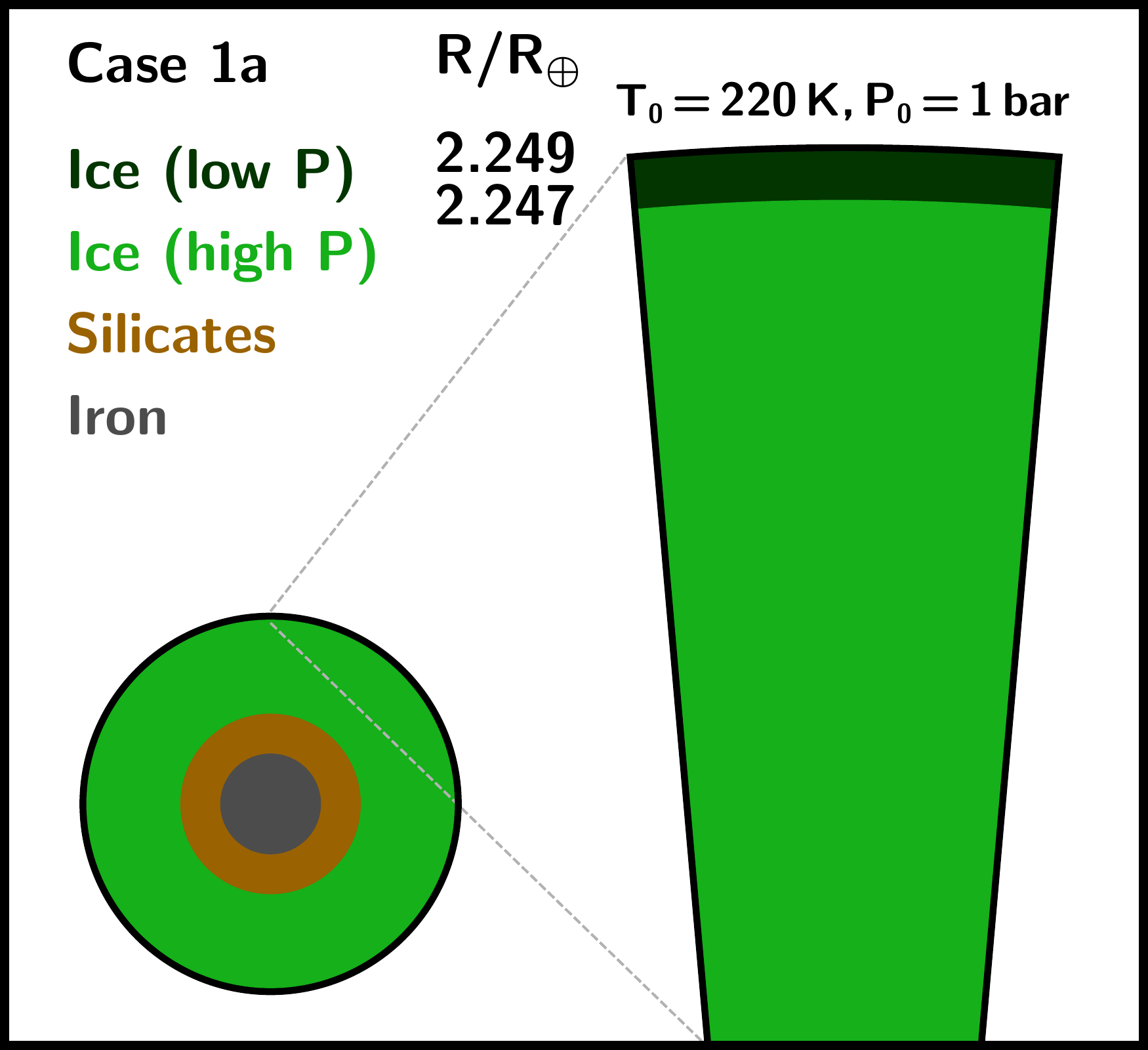}
\includegraphics[width=0.325\textwidth]{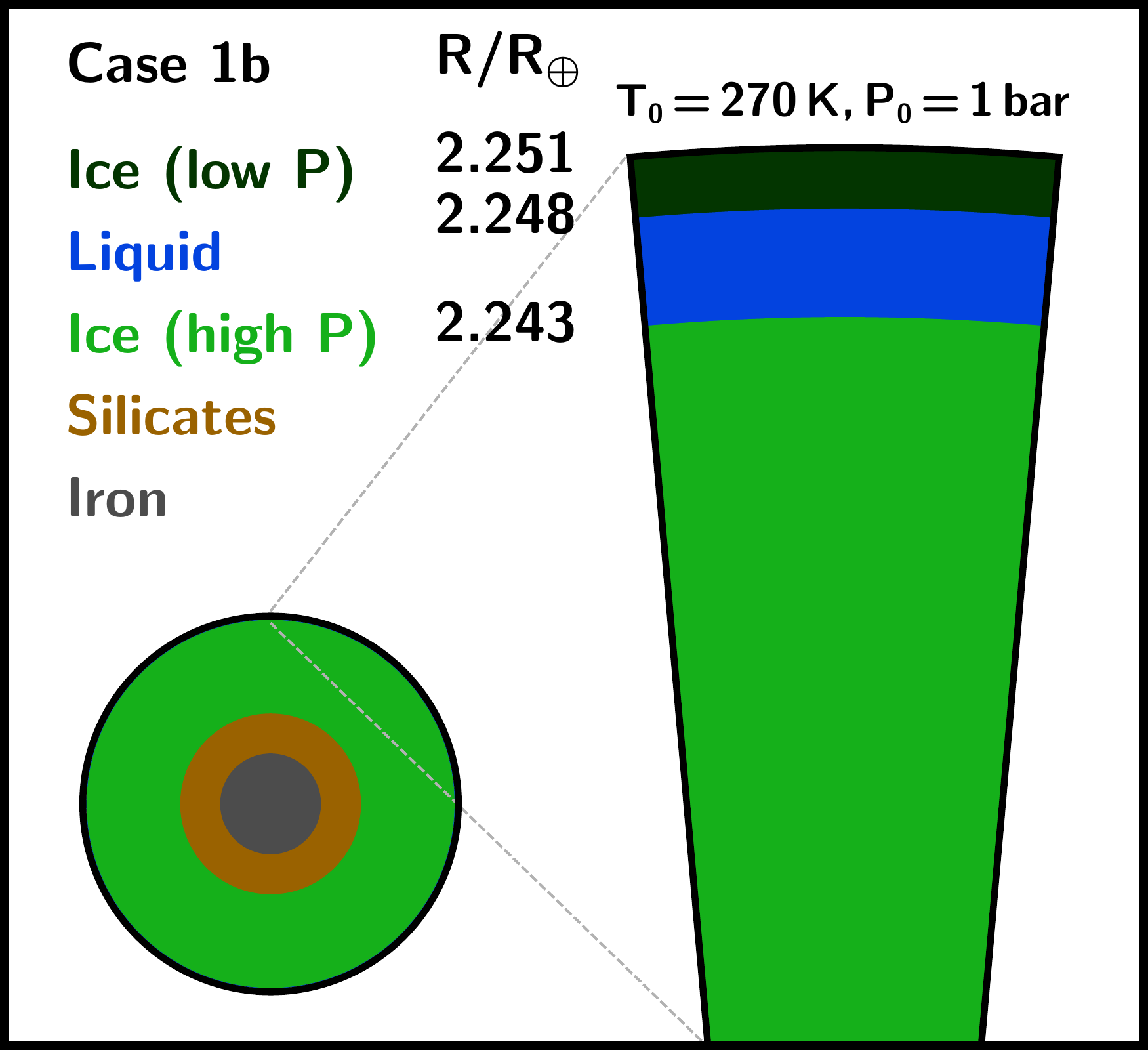} \\
\includegraphics[width=0.325\textwidth]{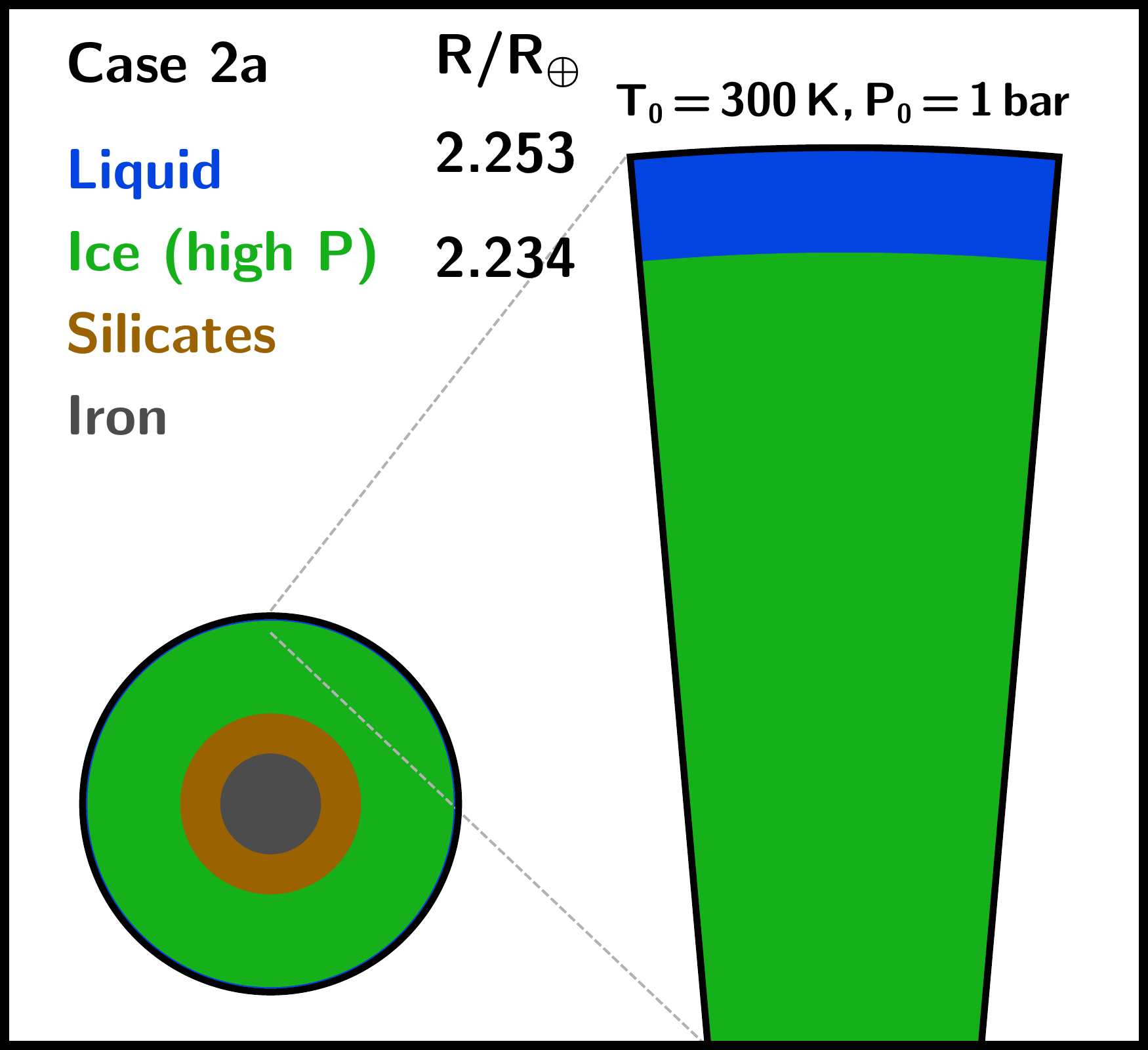} 
\includegraphics[width=0.325\textwidth]{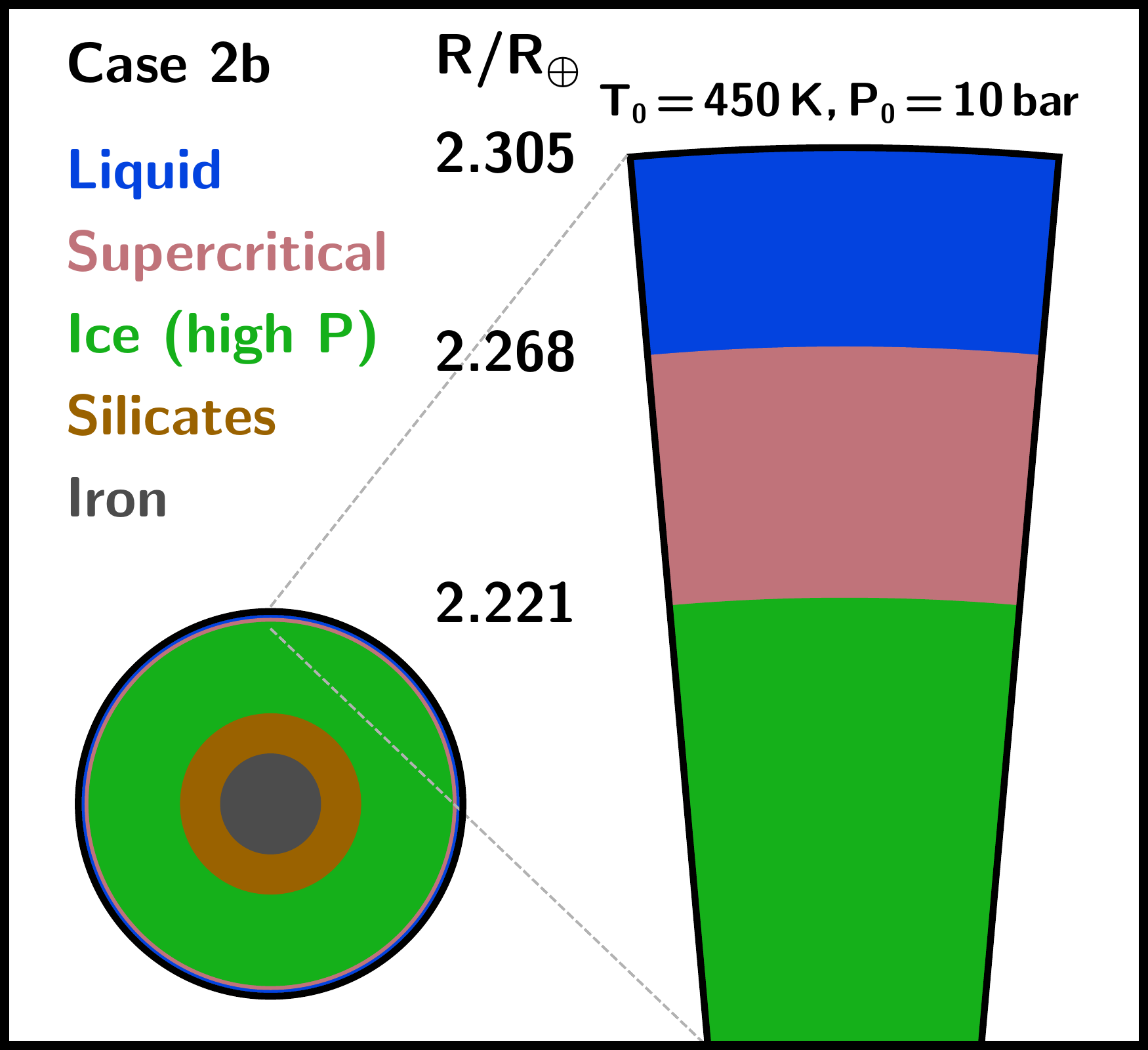} \\ 
\includegraphics[width=0.325\textwidth]{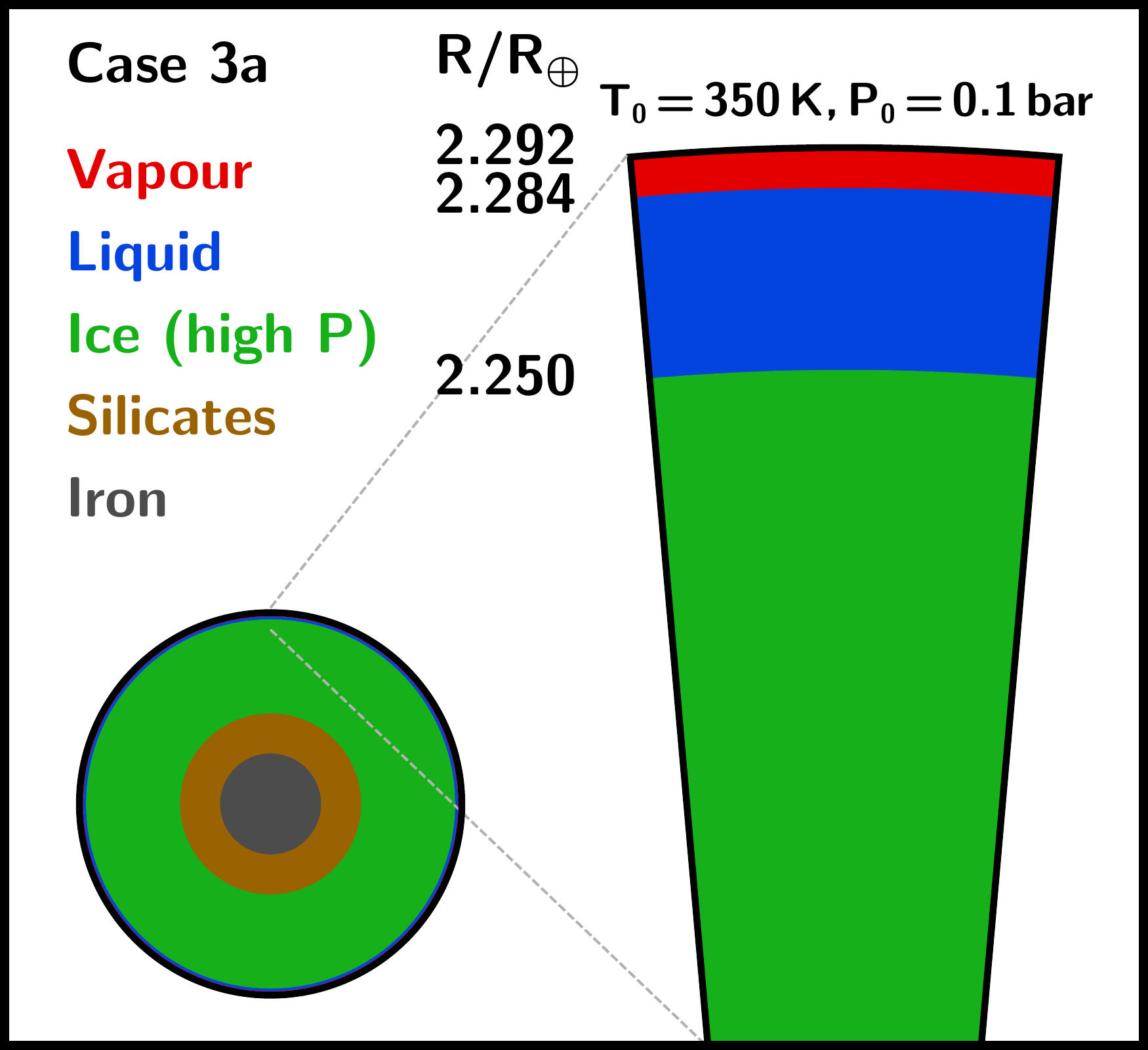}
\includegraphics[width=0.325\textwidth]{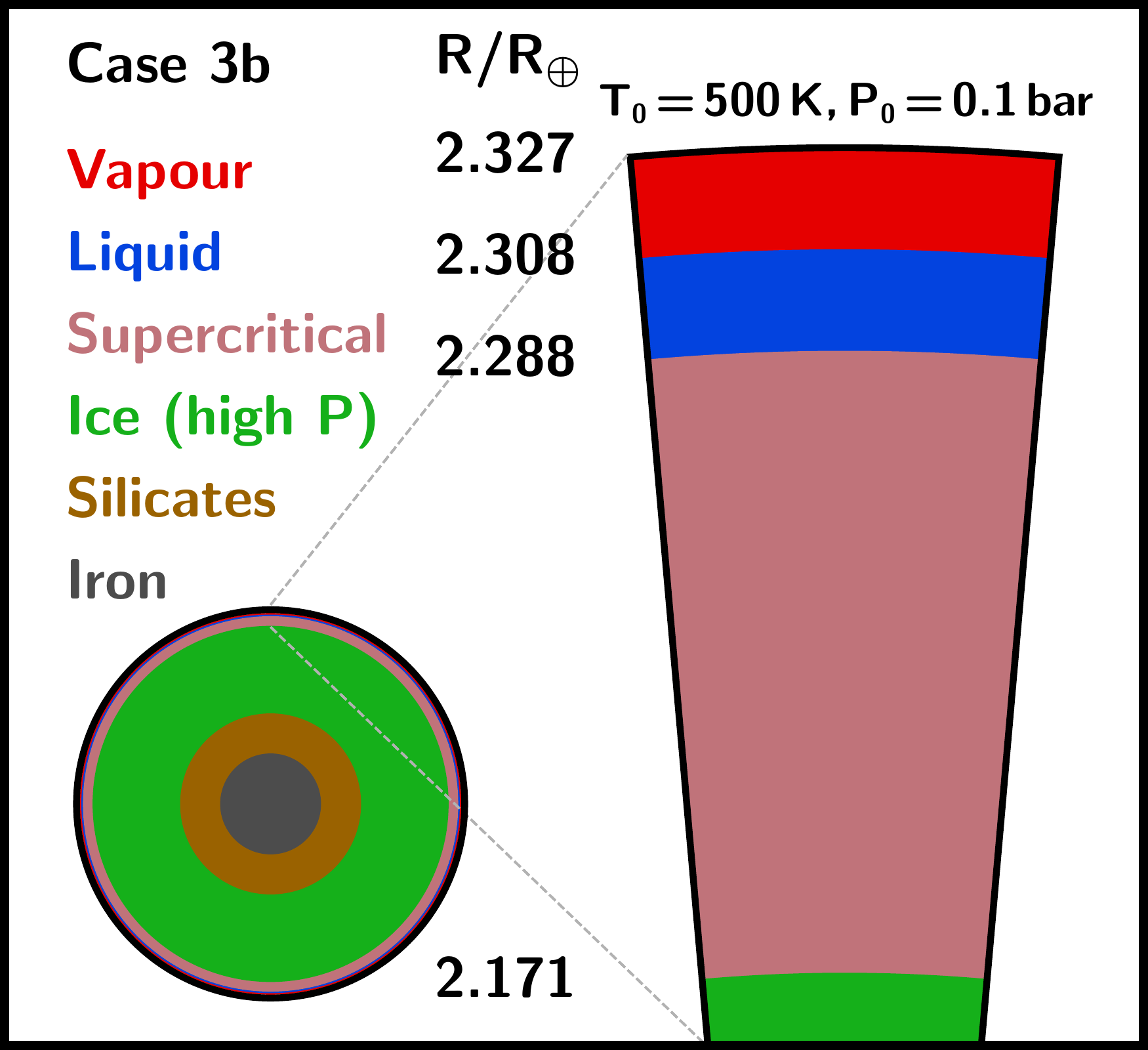} 
\includegraphics[width=0.325\textwidth]{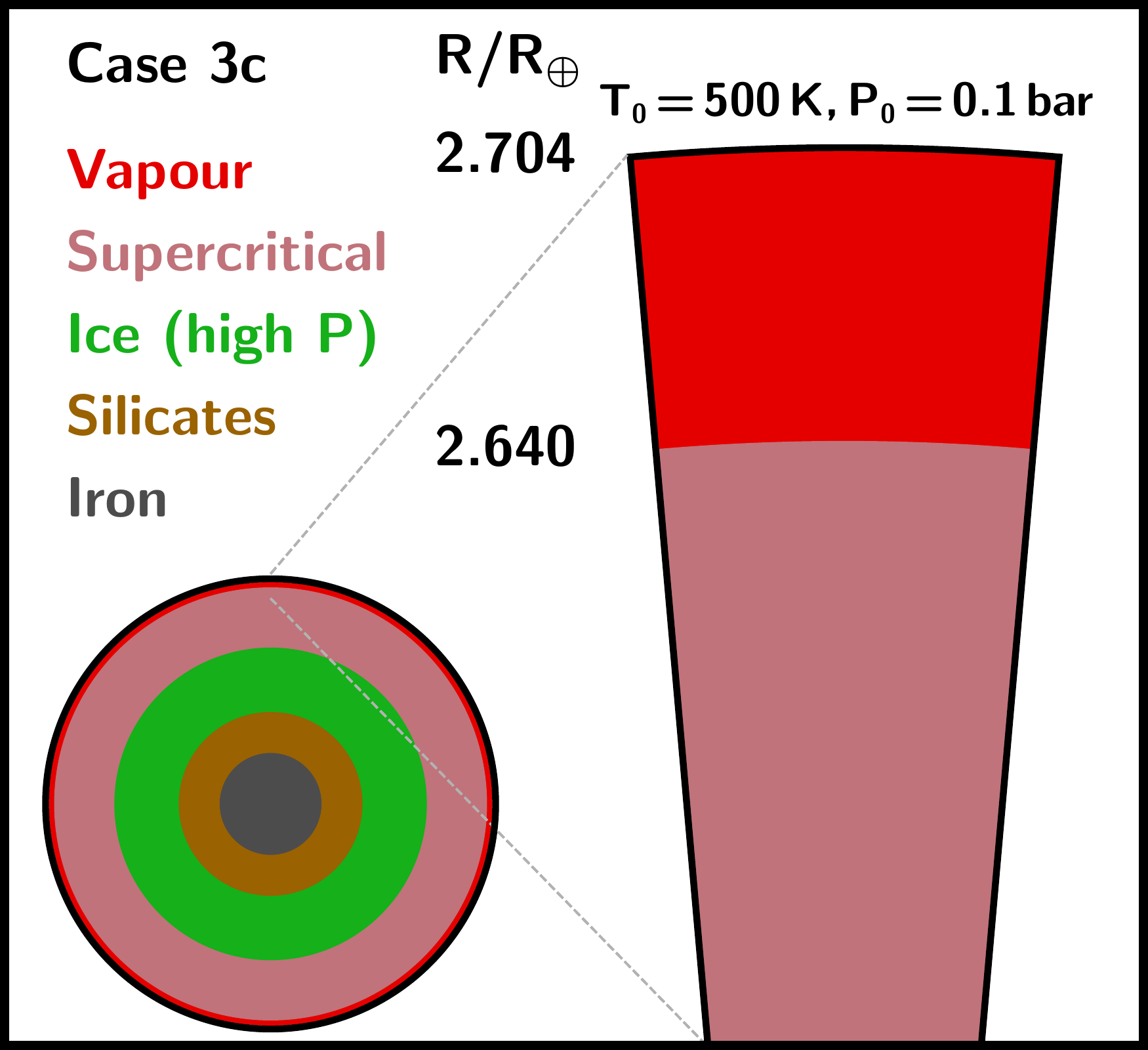}
\end{array}$
\end{center}
    \caption{Phase structures of H$_2$O-rich planets with different surface conditions and temperature structures. Each of these planets has the same mass ($8M_{\oplus}$) and H$_2$O mass fraction (0.7), but different surface conditions. The $P$--$T$ profiles of the H$_2$O layers of these planets are shown in Figure \ref{fig:water_phases}. We find that a diverse set of phase structures are possible depending on the surface conditions and temperature profile, from a sub-surface ocean between low- and high-pressure ice (Case 1b) to a steam atmosphere above a layer of supercritical water (Case 3c).}
    \label{fig:h2o_cases}
\end{figure*}

For illustration, we consider planets of 8$M_{\oplus}$ consisting of a 70\% H$_2$O layer by mass over an Earth-like nucleus (1/3 Fe, 2/3 MgSiO$_3$). We exhaustively search the parameter space of surface conditions to identify the various phase structures that can arise in water world interiors, and choose a representative case for each unique structure that we find. Each of these cases has a surface in one of three phases: ice, liquid or vapour. In the remainder of this section we discuss the structure for each case in turn. Table \ref{tab:boundary_conditions} and Figure \ref{fig:water_phases} summarise the temperature structures of our representative model planets, and Figure \ref{fig:h2o_cases} shows the resulting phase structures for each of these cases. In this figure we refer to ice Ih as "low-pressure ice" and other ice phases as "high-pressure ice". The various cases are now discussed in more detail.

\subsubsection*{Case 1: Ice surface}

Planets with ice Ih surfaces may have one of two different phase structures: they can remain in ice throughout the interior (Case 1a) or host a liquid layer in between low- and high-pressure ices (Case 1b). A completely icy structure prevails in model planets with $T_{0} < 251.2\,$K. This is represented by Case 1a, which has a surface temperature of 220$\,$K and surface pressure of 1$\,$bar. The interior passes through ices Ih, III, V, VI, VII and X. However, a hotter surface temperature that is closer to the ice-liquid phase boundary yields a sub-surface ocean in between ices Ih and the high-pressure ices (in this case ices VI, VII and X). This is demonstrated with Case 1b, for which the model inputs are identical to 1a except for a higher $T_{0}$ of 270$\,$K. A structure similar to Case 1b has been proposed previously for water-rich exoplanets at large orbital separations \citep{Ehrenreich2006}. Allowing for deviations from an adiabatic temperature profile by considering processes such as conduction in the ice layers could heat the interior further, increasing the probability of a subsurface ocean.

\subsubsection*{Case 2: Liquid surface}

Water-rich planets with a liquid surface host predominantly icy H$_2$O layers, either moving directly from liquid to high-pressure ice (Case 2a) or hosting a supercritical layer in between (Case 2b). For most surface conditions, the H$_2$O layer transitions directly from liquid to ices VII and X, as demonstrated by Case 2a which has $T_{0} = 300\,$K and $P_{0} = 1\,$bar. In a small number of cases where $T_{0}$ and $P_{0}$ are large it is possible to obtain a layer of supercritical H$_2$O between the liquid and ice phases. This is shown in Case 2b which has $T_{0} = 450\,$K, $P_{0} = 10\,$bar.

\subsubsection*{Case 3: Steam atmosphere}

For planets whose photospheric pressure and temperature correspond to the vapour phase, leading to a steam atmosphere, the underlying phase structure depends strongly on the atmospheric temperature profile, with interiors consisting of liquid water (Case 3a), supercritical water (Case 3c), or both (Case 3b) before transitioning to high-pressure ice. In order to demonstrate this we consider end-member cases of a purely isothermal and purely adiabatic profile, each with the same photospheric pressure, $P_{0}=0.1\,$bar. For a planet to host a surface ocean, the temperature throughout the envelope must remain below the critical temperature of water, 647$\,$K. Cases 3a and 3b demonstrate this for two different values of $T_0$: in Case 3a, which has a lower $T_{0}$ of 350$\,$K, the interior moves directly from liquid to high-pressure ice, giving a similar structure to Case 2a underneath the atmosphere. By contrast, the hotter $T_0=500\,$K used for Case 3b yields an additional layer of supercritical H$_2$O, which is similar to Case 2b. Finally, if an adiabatic temperature profile in the atmosphere is assumed, the vapour-liquid transition is forbidden and so the interior changes directly from vapour to supercritical H$_2$O, as shown in Case 3c. This is also the case for any planet with $T_{0} \geq 647\,$K. Note that this also inflates the radius of the planet significantly.

\begin{table}
	\centering
	\caption{Surface conditions and radiative convective boundaries for the model planets considered in Section \ref{subsec:h2o_phase}. Each model planet represents a unique phase structure that a water-rich sub-Neptune may possess.}
	\hfill \\
	\label{tab:boundary_conditions}
	\setlength{\arrayrulewidth}{1.3pt}
	\begin{tabular}{ccccc}
		\hline
		Case & Phase at $R_p$ & $T_0$ (K) & $\log P_0$ (bar) & $\log P_{\rm rc}$ (bar) \\
		\hline
		1a & Ice & 220 & 0 & 0 \\
		1b & Ice & 270 & 0 & 0 \\
		2a & Liquid & 300 & 0 & 0 \\
		2b & Liquid & 450 & 1 & 1 \\
		3a & Vapour & 350 & -1 & -0.3 \\
		3b & Vapour & 500 & -1 & 1.5 \\
		3c & Vapour & 500 & -1 & -1 \\
		\hline
	\end{tabular}
\end{table}

\subsection{Planets with mixed envelopes}
\label{subsection:mix}

So far we have considered models where each layer in the planetary interior is fully differentiated. However, it is possible that sub-Neptunes could host envelopes consisting of mixed H$_2$O and H/He, as has been suggested for the interiors of giant planets \citep{Soubiran2015}. H$_2$O is regularly detected in exoplanet atmospheres, and mass--metallicity trends derived from atmospheric observations suggest that lower-mass planets should have higher H$_2$O abundances \citep{Welbanks2019}. In this section we aim to determine the difference between $M$--$R$ relations for mixed and unmixed H$_2$O-H/He envelopes, where the envelope is defined here to mean the outer H$_2$O and H/He components of the planet. This will indicate whether interior structure models need to take into account the presence of atmospheric species as well as highlighting possible degeneracies between mixed and unmixed envelopes of different compositions. We produce $M$--$R$ relations for mixed H$_2$O-H/He envelopes with different quantities of each component, and compare these to the equivalent unmixed relations. The methods for modelling mixed envelopes are described in Section \ref{section:methods}. The helium mass fraction within the H/He component is held constant throughout ($Y=0.275$).

For all models considered here we take $T_0=500\,$K, $P_0=0.1\,$bar and $P_{\rm rc}=10\,$bar. This ensures that the temperature profile is sufficiently hot for the H$_2$O to be in vapour or supercritical phase throughout the envelope, where miscibility with H/He is more likely. The model planets consist of an Earth-like nucleus with a mass fraction of 0.95. The remaining 5\% of the planetary mass is divided between H$_2$O and H/He in different proportions, ranging from a pure H$_2$O to a pure H/He envelope. For each composition we produce $M$--$R$ relations for both a mixed and an unmixed envelope.

\begin{figure}
\includegraphics[width=\columnwidth]{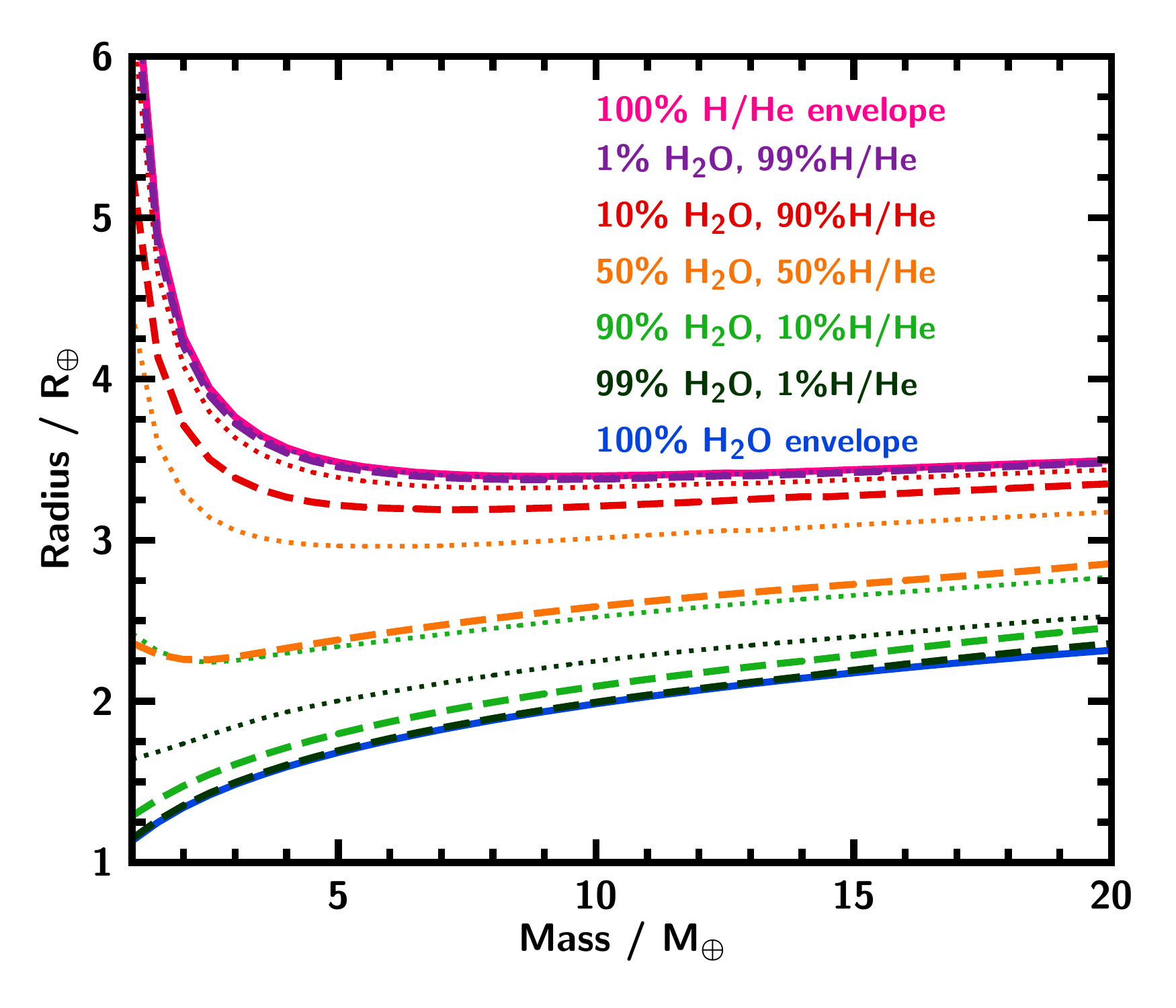}
    \caption{$M$--$R$ relations for planets with mixed (dashed) and unmixed (dotted) envelopes of fixed mass fraction ($x_{\rm env}=0.05$) above an Earth-like nucleus. All models have $T_0=500\,$K, $P_0=0.1\,$bar and $P_{\rm rc}=10\,$bar. The solid blue and pink lines show end-member cases of a pure H$_2$O and H/He envelope respectively. For H/He-rich envelopes with small amounts of H$_2$O, there is little difference between the mixed and unmixed cases. However, for planets with mostly H$_2$O envelopes, mixing in the envelope leads to a much smaller radius than the unmixed case.}
    \label{fig:mixed_env}
\end{figure}

The resulting $M$--$R$ relations are shown in Figure \ref{fig:mixed_env}. We find that small amounts of H$_2$O in a H/He-rich envelope do not significantly alter a planet's radius. A planet with a mixed envelope containing 1\% H$_2$O by mass has a radius that is on average $0.028R_{\oplus}$ lower than a planet of the same total mass with a pure H/He atmosphere, which is small compared to the measurement uncertainties of sub-Neptune radii. An unmixed envelope with 1\% H$_2$O lies in between these cases. This mass fraction corresponds to a volume mixing ratio of $\log X_{\rm H_2O} \approx -2.9$, close to solar H$_2$O abundance at this temperature, $\log X_{\rm H_2O} = -3.0$ \citep{Asplund2009,Madhu2012_CtoO}, and suggests that when modelling the interior structures of planets with hydrogen-dominated atmospheres that may contain small amounts of other chemical species, the assumption of a pure H/He outer envelope provides a sufficient density profile. Conversely, while the difference between a pure H$_2$O envelope and a mixed envelope with 1\% H/He is also minimal, the radius increases substantially when considering an unmixed rather than a mixed envelope. In this case, switching from a mixed to unmixed envelope drastically changes the planet's structure: a mixed envelope has a mostly steam atmosphere throughout, whereas the unmixed case with $T_0=500\,$K has a H/He atmosphere over a layer of supercritical water. It is worth noting that the $M$--$R$ relation for a planet with a mixed envelope consisting of equal amounts of H/He and H$_2$O is very close to the relation for an unmixed envelope with 90\% H$_2$O, 10\% H/He. This provides another source of uncertainty when constraining a planet's composition and interior structure. However, this can potentially be resolved using atmospheric observations.

\section{Summary and Discussion} \label{section:discussion}

We have investigated the internal structures of water-rich exoplanets, motivated by the recent suggestion of the possibility of a habitable water ocean under a H/He atmosphere in the mini-Neptune K2-18b \citep{Madhu2020}. We have presented a modelling framework for sub-Neptune planets, spanning super-Earths and mini-Neptunes, and used this to explore the H$_2$O phase structures of such planets in detail. We have found the following key results:

\begin{itemize}
    \item For planets with a liquid H$_2$O layer, the depth of the ocean is determined by the planet's surface gravity and temperature, and can extend up to $\sim$1600$\,$km. Planets with gravities comparable to Earth's can have oceans that are over one hundred times deeper than the Earth's ocean, which has an average depth of $\sim$3.7$\,$km.
    \item Planets with H/He envelopes can allow for significant liquid H$_2$O layers underneath the envelope. This phenomenon can occur over a wide region in $M$--$R$ space. For example, liquid H$_2$O may exist on such planets with $T_0=300\,$K, $P_{\rm rc}=10\,$bar provided the total mass of the H/He envelope is $\lesssim 4 \times 10^{-4}M_{\oplus}$. The mass of H/He permitted increases as the photospheric temperature decreases and as the radiative-convective boundary moves deeper in the atmosphere.
    \item More broadly, a diverse range of phase structures are possible in H$_2$O-rich super-Earths and mini-Neptunes. Besides surface liquid H$_2$O, other phase structures include liquid water sandwiched between two ice layers or a steam atmosphere above supercritical water and high-pressure ice, depending on surface conditions.
    \item Planets with mixed H/He-H$_2$O envelopes have significantly lower radii than planets with the same composition but differentiated H/He and H$_2$O if the mass fractions of the two components are comparable, but this effect is minimal for a small mass fraction of H$_2$O ($\sim$1$\%$) in a mostly H/He atmosphere.
\end{itemize}

\subsection{Potential habitability of water worlds}

Traditionally, a planet is classed as habitable if it has the right conditions for liquid water at its surface \citep{Hart1979,Kasting1993}. Living organisms have been found to survive in liquid water at $T\lesssim400\,$K and $P\lesssim1000\,$bar \citep{Merino2019}. Some of the possible interior structures found for K2-18b in \citet{Madhu2020} had thermodynamic conditions in this range at the surface of the water layer. Many of the model planets considered in this work also fit this definition, highlighting the possibility of a family of habitable planets in the mini-Neptune regime.

It is important to consider whether other factors could preclude habitability for water worlds. All of the planets with H$_2$O oceans shown in Section \ref{subsec:h2o_phase} have very deep layers of high-pressure ice which separate the ocean from the silicate mantle. \citet{Kitzmann2015} found that this separation, which prevents a carbonate-silicate cycle that can regulate the CO$_2$ inventory of a planet, can have a destabilising effect on the climate with negative consequences for habitability. However, other CO$_2$ exchange mechanisms have been proposed that may have a regulatory effect \citep{Levi2017,Ramirez2018}. The potential for habitable conditions on H$_2$O-rich planets should therefore not be ruled out despite the lack of an Earth-like carbonate-silicate cycle.

In Section \ref{subsection:ocean} we consider ocean depths for surface temperatures ranging from 273--584$\,$K. Past studies have been restricted to lower temperatures; for example \citet{Noack2016} explores temperatures from 290--370$\,$K, stating that for higher temperatures the liquid water in the ocean would evaporate due to the runaway greenhouse effect. However, this rests on the assumption of a surface pressure of 1$\,$bar similar to Earth.  By assuming a surface pressure of 100$\,$bar we can explore higher temperatures, and we find that the increase in ocean depth with surface temperature reported by \citet{Noack2016} no longer holds beyond $T_0=413\,$K. The pressure at the surface of the ocean on a water world is not known and could be greater than 1$\,$bar, especially if the planet possesses a H/He envelope. Relaxing this assumption allows for a widening of the parameter space for water worlds with global oceans and suggests that liquid water layers could be a possibility even for planets considerably warmer than Earth at the bottom of their atmospheres.

In Figure \ref{fig:liquid_mr} we show a region of $M$--$R$ space where planets require H/He envelopes but could still host liquid water oceans. This region is found assuming a maximal H$_2$O mass fraction of 90\%. While it is understood that a minimal amount of rocky material is necessary to form a planet by core accretion, the exact range of possible ice/rock ratios for exoplanets is not known, and estimates vary widely. It is possible that a planet consisting of as much as 90\% H$_2$O and only $\sim$10$\%$ iron and silicates could be unfeasible from a planet formation perspective. If this is the case, a region of $M$--$R$ space similar to the one described here will still exist, but at smaller radii due to the lower maximum radius of a planet without any H/He. Whether planets that fall into this category are indeed habitable depends on what effect a hydrogen-rich atmosphere has on a planet's climate, which is still poorly understood. However, a number of studies have argued in favour of rocky planets with H/He atmospheres being potentially habitable \citep{Pierrehumbert2011,Koll2019} and other works \citep[e.g.,][]{Seager2020} have shown that single-celled organisms can survive in a pure H$_2$ atmosphere. The potential for habitability of sub-Neptunes with H$_2$-rich atmospheres is considered in several recent and upcoming works \citep[][Madhusudhan et al., in press]{Madhu2020,Piette2020}.

\subsection{Future directions and applications}

There are many challenges to overcome in order to characterise water-rich exoplanets in detail, both from a theoretical and observational perspective. Our model makes a number of standard assumptions that may be challenged in the future. For example, while the vast majority of internal structure models assume an adiabatic temperature profile throughout most of the interior, thermal boundary layers that inhibit convection have been proposed for the interiors of Uranus and Neptune \citep{Nettelmann2016,Podolak2019}. Additionally, while we consider mixed H/He-H$_2$O envelopes in this study, other chemical species will be present in planetary interiors that might affect their internal structure. \citet{Levi2014} modelled water-rich planets with a methane component, and while they found that the inclusion of methane did not significantly change the $M$--$R$ relation, they did see a noticeable effect on the atmosphere. \citet{Shah2020} modelled planets with hydrated interiors, and while the impact on $M$--$R$ relationships was small compared to current measurement uncertainties, future missions such as PLATO \citep{Rauer2016} may lead to a higher precision for masses and radii that could make these effects important to consider. Many aspects of the behaviour of high-pressure ices, such as possible interactions between rock and ice in this regime, are still unknown \citep{Journaux2020,Huang2021}, and further understanding in this area could have important consequences for the potential habitability of ocean worlds \citep{Noack2016}.

Characterising observed sub-Neptunes using internal structure models is an inherently degenerate problem, with many different compositions consistent with a given mass and radius. For many planets in this regime, it is impossible to distinguish between a water world scenario and a rocky planet with a thick H/He envelope \citep[e.g.][]{Luque2021}. Determining the atmospheric compositions of these planets may allow us to break this degeneracy: while a small number of mini-Neptune atmospheres have been characterised using spectroscopic observations \citep[e.g.,][]{Benneke2019}, it is hoped that upcoming facilities such as the \textit{James Webb Space Telescope} will provide the opportunity to observe the atmospheres of low-mass planets in much greater detail than has previously been possible \citep[e.g.,][]{Morley2017,Welbanks2021}, which should in turn allow better constraints on the interior composition and structure.

The results from this study, as well as those of \citet{Madhu2020} and \citet{Piette2020}, highlight the diversity of exoplanets that have the potential to be habitable. These results point to the exciting possibility of the right conditions for life being present on planets much larger than Earth. As next-generation instruments make potential ocean planets more amenable to characterisation and facilitate the search for biosignatures, we hope that our findings can further motivate the quest to detect signs of life on other worlds, even those which bear little resemblance to our own.

\section*{Acknowledgements}
MN acknowledges support from the Science and Technology Facilities Council (STFC), UK, towards his PhD programme. The authors thank Anjali Piette for providing model temperature profiles for comparison as shown in Figure \ref{fig:k2_pt_check} and the anonymous reviewer for helpful comments on the manuscript. This research has made use of the NASA Astrophysics Data System and the Python packages \textsc{numpy}, \textsc{scipy} and \textsc{matplotlib}.

%%%%%%%%%%%%%%%%%%%%%%%%%%%%%%%%%%%%%%%%%%%%%%%%%%

\section*{Data Availability}
No new data is generated in this work. 


\bibliographystyle{mnras}
\bibliography{main}

\appendix

\section{Mass--radius relations for water worlds} \label{appendix:a}

Here we describe in more detail how variations in a number of different parameters can affect $M$--$R$ relations for planets with and without H/He envelopes.

\subsection*{Water worlds without H/He envelopes}

It is clear from Figure \ref{fig:h2o_summary} that variations in $T_0$ can have a significant effect on planetary radius in both the isothermal and adiabatic cases. For model scenarios with a large H$_2$O mass fraction of 0.9 and an isothermal vapour layer, increasing $T_0$ from $300\,$K to $1000\,$K inflates the planetary radius by $0.25R_{\oplus}$--$0.4R_{\oplus}$ depending on the planet mass. If the vapour layer is adiabatic the radius increases more dramatically, by up to $2.2R_{\oplus}$ in the most extreme case. The inflation is enhanced at lower masses due to the low surface gravity of the planet allowing for a more extended envelope \citep{Rogers2011}. This effect is still significant for planets with a lower H$_2$O mass fraction: even in the isothermal vapour case, at $x_{\rm H_2O}=0.3$ the radius is inflated by $0.1-0.2R_{\oplus}$ for the $1000\,$K model planets compared to those at $300\,$K. We note that even in the case of a minimal H$_2$O layer, the effect of changing $T_0$ may be non-negligible: for $x_{\rm H_2O}=0.01$, the radius of a $1M_{\oplus}$ planet with $T_0=300\,$K is $0.981R_{\oplus}$, increasing to $1.076R_{\oplus}$ at $T_0=1000\,$K with an isothermal vapour layer or $1.171R_{\oplus}$ with an adiabatic vapour layer. Even for mostly rocky planets that may host thin H$_2$O envelopes, temperature dependence within the H$_2$O layer is still important to consider.

Figure \ref{fig:h2o_summary} also shows how $P_0$ affects the planetary radius. In general, a higher $P_0$ leads to a smaller radius if all other parameters are left unchanged. If an isothermal temperature profile for vapour is assumed, this change is quite small: $\lesssim 0.001R_{\oplus}$ for planets with $T_0=300\,$K, and approaching $0.1 R_{\oplus}$ for the lowest masses ($M \sim 1M_{\oplus}$) at $T_0=1000\,$K. However, the choice of $P_0$ affects the $M$--$R$ relation more significantly if the temperature profile in the vapour layer is adiabatic. At $10M_{\oplus}$, the radius of a 90\% water planet with $T_0=1000\,$K increases by 0.47$R_{\oplus}$ if $P_0$ changes from 0.1--100$\,$bar. The effect is more pronounced at higher temperatures since the density of vapour and supercritical H$_2$O varies more strongly with pressure than the density of liquid (see Figure \ref{fig:h2o_eos}). If the vapour layer is adiabatic, then the temperature profile increasing from a lower $P_0$ leads to a hotter H$_2$O layer throughout which further magnifies this effect.

We find that a $20M_{\oplus}$, 90\% H$_2$O planet with a solid or liquid photosphere can have a radius of up to $3R_{\oplus}$, in agreement with the findings of \citet{Zeng2019} that planets $>3R_{\oplus}$ generally require a gaseous envelope. Planets with $x_{\rm H_2O} \gtrsim 0.9$ are likely unrealistic from a planet formation perspective, since some amount of refractory material is required to initiate ice and gas accretion \citep{Zeng2014,Lee2016}. 

The $M$--$R$ relations reported in \citet{Mousis2020} yield larger radii than our model for a similar composition. This a result of different methods being used to compute the temperature profile in the water layer. In order to compare the models more directly we consider a 15$M_{\oplus}$ planet with a 20\% H$_2$O layer above a silicate mantle and an adiabatic steam atmosphere with $P_0=0.1\,$bar, $T_0=430\,$K, which gives a temperature profile close to that of \citet{Mousis2020} with $T_{\rm eq}=300$\,K (see figure 1 of that paper). The radius of this model planet is 2.6$R_{\oplus}$, close to the 2.7$R_{\oplus}$ shown in figure 2 of \citet{Mousis2020}.

\subsection*{Water worlds with H/He envelopes}

 Figure \ref{fig:hhe_summary} shows $M$--$R$ relations for a number of water-rich planets with H/He envelopes. We show two values of $T_0$, 300$\,$K and 500$\,$K. We also consider several different radiative-convective boundaries, with $P_{\rm rc}$ ranging from 1--1000$\,$bar (see Section \ref{subsubsec:ia}). The results indicate that only a small amount of H/He by mass is required to substantially increase the radius of a planet. Planets with $T_0>500\,$K may have larger radii than Neptune ($3.88R_{\oplus}$), even with a H/He mass fraction as low as 1\%. Changing $P_{\rm rc}$ also has a very significant effect on the $M$--$R$ relation. This highlights the importance of using an accurate atmospheric temperature profile when modelling planets with a gaseous envelope. In cases where a planet's temperature profile is not well-constrained, a wide variety of $P$--$T$ profiles should be considered in order to fully explore its possible internal structures. These thermal effects become more significant as the envelope mass increases.

Throughout this section we assume a nominal photospheric pressure of 0.1$\,$bar. Analysis of the transmission spectra of hydrogen-rich planets yield values of $P_0$ ranging from $\sim$1$\,$mbar--$\sim$1$\,$bar \citep[e.g.][]{Welbanks2019}. It is straightforward to compute the change in radius resulting from a change in $P_0$, since at pressures below $\sim$1000$\,$bar the EOS for the H/He envelope is that of an ideal gas. Combining this with equation \ref{eq:hydro_eqm} shows that the change in adjusting the pressure at the photosphere from $P_1$ to $P_2$ yields a change in radius of $NH_{\rm sc}$, where $N = \ln (P_1/P_2)$ and $H_{\rm sc}=k_BT/\mu g$ is the atmospheric scale height. Adjusting $P_0$ from 1$\,$bar to 1$\,$mbar increases the planetary radius by $\sim$7$\,H_{\rm sc}$.

% Don't change these lines
\bsp	% typesetting comment
\label{lastpage}
\end{document}